\documentclass[10pt]{iopart}

\expandafter\let\csname equation*\endcsname\relax 
\expandafter\let\csname endequation*\endcsname\relax
\usepackage{amsmath}
\usepackage{amssymb}
\usepackage{graphicx}
\usepackage{newtxtext}
\usepackage[varvw]{newtxmath}
\usepackage{bbm}
\usepackage{physics}
\usepackage{slashed}
\usepackage{tabularx}
\newcolumntype{Y}{>{\centering\arraybackslash}X}
\usepackage[font=small]{caption}
\usepackage{subcaption}
\usepackage{mathtools}
\usepackage{ragged2e} 
\usepackage{chngcntr} 
\usepackage{setspace}

\usepackage{hyperref}

\usepackage[square, numbers, sort&compress]{natbib} 

\makeatletter
\newcommand{\mainmatter}{%
  \setcounter{footnote}{0}
  \patchcmd{\@makefntext}{\fnsymbol}{\arabic}{}{}
  \patchcmd{\@thefnmark}{\fnsymbol}{\arabic}{}{}
  \def\@makefnmark{\textsuperscript{\arabic{footnote}}}
  \def\@makefntext{\textsuperscript{\arabic{footnote}}}
}
\makeatother

\hypersetup{
    colorlinks,
    linkcolor={blue},
    citecolor={blue},
    urlcolor={blue}
}

\graphicspath{{./}{./figs/}}

\newcommand{\Nc}{N_\mathrm{c}}
\newcommand{\llangle}{\langle\!\langle}
\newcommand{\rrangle}{\rangle\!\rangle}
\newcommand{\lllangle}{\langle\!\langle\!\langle}
\newcommand{\rrrangle}{\rangle\!\rangle\!\rangle}

\DeclareMathAlphabet{\mymathbb}{U}{BOONDOX-ds}{m}{n}

\makeatletter
\let\MyIntOrig\int
\def\MyIntSpace{\hspace{-.25em}} 
\def\int{\MyInt}
\def\MyInt{\MyIntOrig\MyIntSkipMaybe}
\def\MyIntSkipMaybe{
  \@ifnextchar_{\MyIntSkipScript}{%
  \@ifnextchar^{\MyIntSkipScript}{%
  \@ifnextchar\limits{\MyIntSkipTok}{%
  \@ifnextchar\nolimits{\MyIntSkipTok}{%
  \MyIntSpace}}}}%
}
\def\MyIntSkipScript#1#2{#1{#2}\MyIntSkipMaybe}
\def\MyIntSkipTok#1{#1\MyIntSkipMaybe}
\makeatother

\makeatletter
\newcommand{\vast}{\bBigg@{3.5}}
\newcommand{\Vast}{\bBigg@{4}}
\makeatother

\DeclareCaptionJustification{justified}{\justifying}

\date{\today}

\usepackage[dvipsnames]{xcolor}
\usepackage[normalem]{ulem}

\begin{document}

\submitto{\RPP}

\title{Continuous order-to-order quantum phase transitions from fixed-point annihilation}

\author{David J. Moser and
Lukas Janssen}

\address{Institut f\"ur Theoretische Physik and W\"urzburg-Dresden Cluster of Excellence {\it ct.qmat}, TU Dresden, 01062 Dresden, Germany}

\begin{abstract}
A central concept in the theory of phase transitions beyond the Landau-Ginzburg-Wilson paradigm is fractionalization: the formation of new quasiparticles that interact via emergent gauge fields.
This concept has been extensively explored in the context of continuous quantum phase transitions between distinct orders that break different symmetries.
We propose a mechanism for continuous order-to-order quantum phase transitions that operates independently of fractionalization.
This mechanism is based on the collision and annihilation of two renormalization group fixed points: a quantum critical fixed point and an infrared stable fixed point.
The annihilation of these fixed points rearranges the flow topology, eliminating the disordered phase associated with the infrared stable fixed point and promoting a second critical fixed point, unaffected by the collision, to a quantum critical point between distinct orders.
We argue that this mechanism is relevant to a broad spectrum of physical systems. 
In particular, it can manifest in Luttinger fermion systems in three spatial dimensions, leading to a continuous quantum phase transition between an antiferromagnetic Weyl semimetal state, which breaks time-reversal symmetry, and a nematic topological insulator, characterized by broken lattice rotational symmetry. 
This continuous antiferromagnetic-Weyl-to-nematic-insulator transition might be observed in rare-earth pyrochlore iridates $R_2$Ir$_2$O$_7$.
Other possible realizations include kagome quantum magnets, quantum impurity models, and quantum chromodynamics with supplemental four-fermion interactions.
\end{abstract}

\ioptwocol

\maketitle

\mainmatter

\section{Introduction}
%
The exploration of quantum phase transitions beyond the Landau-Ginzburg-Wilson paradigm~\cite{senthil04a, senthil04b} has garnered significant attention in recent years.
A particular focus has been on continuous quantum phase transitions between phases that break different symmetries~\cite{senthil23}.
Traditionally, it has been believed that such transitions require fractionalization, where new ``deconfined'' quasiparticles with fractional quantum numbers arise and interact via emergent gauge fields.
The intense study of deconfined quantum phase transitions in recent years has significantly advanced our understanding of quantum critical phenomena. Notable insights include the discovery of emergent symmetries~\cite{lou09, nahum15b, deng24, demidio24}, the identification of spectroscopic signatures of fractionalization~\cite{ma18}, the uncovering of unexpected webs of field-theoretical dualities~\cite{wang17}, and the observation of anomalous scaling behaviors in entanglement entropy~\cite{zhao22, song25}.
However, the most recent numerical studies suggest that the deconfined quantum phase transition, at least in its original formulation within SU(2) spin models~\cite{senthil04a, senthil04b, senthil23}, is ultimately first-order, characterized by a finite, though large, correlation length at the transition point~\cite{nahum15a, deng24, takahashi24, zhou24, song25}.

\begin{figure}[b!]
\includegraphics[width=\linewidth]{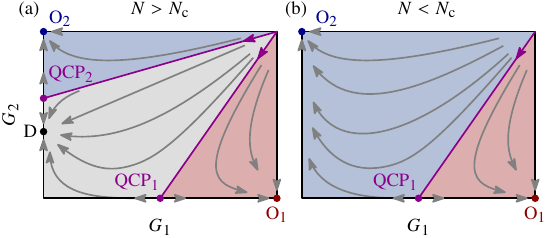}
\caption{%
Mechanism for continuous order-to-order transition from fixed-point annihilation.
(a)~Schematic RG flow in the space spanned by two couplings, $G_1$ and $G_2$, for $N > \Nc$.
Arrows indicate flow towards infrared.
Beyond an interacting disordered phase (gray) at small couplings, governed by the fixed point D, there are two ordered phases (red and blue) in the strong-coupling regime, governed by the fixed points O$_1$ and O$_2$.
The disordered phase is separated from the two ordered phases by continuous phase transition lines (purple), governed by the quantum critical fixed points QCP$_1$ and QCP$_2$.
As $N$ decreases, the fixed points D and QCP$_2$ approach each other and eventually collide at a critical value $\Nc$.
(b)~Same as~(a), but for $N < \Nc$. The fixed-point annihilation at $\Nc$ alters the flow topology, eliminating the disordered phase and leaving behind a single continuous phase transition line (purple) between two phases with distinct orders.
}
\label{fig:intro}
\end{figure}

Here, we propose a mechanism that provides a different pathway to continuous order-to-order quantum phase transitions, one that does not rely on fractionalization. Our proposal involves the collision and annihilation of two renormalization group (RG) fixed points: a quantum critical fixed point and an infrared stable fixed point. We argue that the mechanism is relevant to a broad spectrum of physical systems. 
In particular, we explicitly demonstrate within a RG calculation that three-dimensional Luttinger fermion systems undergo a continuous quantum phase transition between an antiferromagnetic Weyl semimetal state, marked by broken time-reversal symmetry, and a nematic topological insulator state, characterized by broken lattice rotational symmetry. Further possible realizations in other setups are discussed as well.

The mechanism is schematically illustrated in Fig.~\ref{fig:intro}, which shows RG flow diagrams for two different values of an external control parameter $N$.
For $N$ larger than a critical value $\Nc$, there is an interacting disordered phase, which is separated from two ordered phases by continuous phase transition lines, see Fig.~\ref{fig:intro}(a).
As $N$ decreases, the fixed point associated with the disordered phase (D) and those governing one of the two transition lines (QCP$_2$) approach each other and eventually collide at a critical value $\Nc$.
For $N$ smaller than $\Nc$, Fig.~\ref{fig:intro}(b), the fixed-point annihilation alters the flow topology by removing the disordered phase, leaving behind a continuous phase transition line between two phases with distinct orders.
Note that the QCP$_1$ fixed point, which governs the resulting continuous order-to-order transition, is entirely decoupled from the fixed-point annihilation occurring between D and QCP$_2$. 
On the left-hand side of QCP$_1$, for $G_1$ below the fixed-point value, the system would naturally flow toward the disordered phase.
%
%
However, due to the annihilation of the disordered fixed point D with QCP$_2$, this phase becomes unstable, giving rise instead to the competing order.
Notably, since the proposed mechanism solely relies on the fixed-point annihilation of the fixed points D and QCP$_2$ and the resulting rearrangement of the flow topology, any quantum critical fixed point not involved in the annihilation qualifies as QCP$_1$.
%

The remainder of the paper is organized as follows: In Sec.~\ref{sec:pyrochlore-iridates}, we discuss in detail a concrete realization of the proposed mechanism in Luttinger fermion systems in three spatial dimensions, relevant to certain pyrochlore iridates.
Another potential realization, pertinent to kagome quantum magnets, is discussed in Sec.~\ref{sec:kagome}.
We conclude with a summary of results in Sec.~\ref{sec:conclusions}.
The appendices provide technical details of the calculations and discuss additional potential realizations of the mechanism in quantum impurity models and quantum chromodynamics.

\section{Pyrochlore iridates}
\label{sec:pyrochlore-iridates}

\subsection{Model}

We begin by reviewing the interacting Luttinger fermion model in three spatial dimensions~\cite{abrikosov71, abrikosov74, moon13, murakami04}.
This continuum model effectively describes electronic excitations near a quadratic band touching point at the Fermi level, as they occur in pyrochlore iridates $R_2$Ir$_2$O$_7$, with $R$ a rare-earth element~\cite{witczak14,kondo15,nakayama16,armitage18,wang20}.
In the low-temperature limit, the model can be defined by the Euclidean action $S =S_0 + S_\text{int}$, where $S_0$ corresponds to the noninteracting part,
\begin{align}
	S_0 = \int \dd{\tau} \dd[3]{\vec x} \sum_{i=1}^N \psi^{\dagger}_i \left( \partial_\tau + \sum_{a = 1}^5 (1 + s_a \delta) d_a(-\rmi \nabla) \gamma_a \right) \psi_i,
\end{align}
which originates in the Luttinger Hamiltonian~\cite{luttinger56}.
In the above,
$N$ corresponds to the number of band touching points at the Fermi level, with $N=1$ in the case of the pyrochlore iridates,
$\psi_i$ is a four-component Grassmann field, 
$s_a \coloneqq 1$ ($s_a \coloneqq -1$) for $a = 1, 2, 3$ ($a = 4,5$),
$\delta$ parametrizes the cubic anisotropy with $-1 \leq \delta \leq 1$,
the $4 \times 4$ matrices $\gamma_a$ fulfill the Euclidean Clifford algebra, $\{ \gamma_a ,\gamma_b \} = 2 \delta_{ab} \mathbbm 1$,
and $d_1 (\vec{p}) \coloneqq \sqrt{3} p_y p_z$, $d_2 (\vec{p}) \coloneqq \sqrt{3} p_x p_z$, $d_3 (\vec{p}) \coloneqq \sqrt{3} p_x p_y$, $d_4 (\vec{p}) \coloneqq \sqrt{3} (p_x^2 - p_y^2)/2$, and $d_5 (\vec{p}) \coloneqq (2 p_z^2 - p_x^2 - p_y^2)/2$ are proportional to the $\ell = 2$ real spherical harmonics.

We account for both short-range repulsion and the long-range tail of the Coulomb interaction via~\cite{herbut14, janssen15, savary14, janssen16a, janssen17a, boettcher17, moser24}
\begin{align} \label{eq:LagrangianInt}
    S_\text{int} & = 
    - \int \dd{\tau} \dd[3]\vec x \, \biggl\{
    \frac{G_1}{2N} (\psi^\dagger \gamma_{45} \psi )^2
    + \frac{G_2}{2N} \sum_{a=1}^5 (\psi^\dagger \gamma_{a} \psi)^2
    \displaybreak[0] \nonumber \\ & \quad
    + \frac{e^2}{8\pi N} \int \dd[3]{\vec y} \psi^\dagger(\vec{x}) \psi(\vec{x}) \frac{1}{|\vec{x} - \vec{y}|} \psi^\dagger(\vec{y}) \psi(\vec{y}) 
    \biggr\},
\end{align}
where we defined $\gamma_{45} \coloneqq \rmi \gamma_4 \gamma_5$. For notational simplicity, we abbreviated $\psi \equiv \psi_i(\tau,\vec x)$ and $\psi(\vec x) \equiv \psi_i(\tau, \vec x)$, and suppressed the summation symbols over $i=1,\dots,N$.
The long-range tail parameterized by the charge $e^2$ must be taken into account because the Coulomb interaction is only marginally screened due to the vanishing density of states at the Fermi level~\cite{janssen16a}.
The coupling $G_1$ originates from local Hubbard repulsion~\cite{ladovrechis21}, and is believed to increase with decreasing radius of the rare-earth ion~\cite{witczak14}. 
As we show below, a sizable $G_1$ generates a transition towards a Weyl semimetal phase characterized by spontaneous time-reversal symmetry breaking and all-in-all-out (AIAO) antiferromagnetic order on the pyrochlore lattice~\cite{savary14, boettcher17, ladovrechis21, moser24}.
The coupling $G_2$ may be assumed to be small on the microscopic level. However, as will be shown below, $G_2$ is generated by $G_1$ and the long-range interaction, and we therefore include it from the outset.
A sizable $G_2$ drives a transition towards a nematic phase characterized by spontaneous lattice rotational symmetry breaking and a topological gap in the fermionic spectrum~\cite{herbut14, janssen15, janssen17a}.
While previous works have focused exclusively on instabilities toward either AIAO order~\cite{savary14, boettcher17, ladovrechis21, moser24} or nematic order~\cite{herbut14, janssen15, janssen17a}, the present study treats both order parameters on equal footing. This proves essential for revealing the direct order-to-order transition.

\subsection{Hubbard-Stratonovich transformation}
\label{subsec:Hubbard-Stratonovich}

It is convenient to rewrite the interactions in the Luttinger model in terms of Hubbard-Stratonovich fields $a$, $\phi$, and $\varphi$.
In particular, this formulation is crucial in order to explicitly compute order parameters within the RG approach, since it avoids the runaway flows encountered in the purely fermionic formulations~\cite{herbut14,janssen17a,boettcher17}.

We introduce the scalar Coulomb field $a$ via
\begin{align}
    S_a = \int \dd{\tau} \dd[3]{\vec x} \left[
    \frac{1}{2} (\nabla a)^2 + \frac{\rmi e}{\sqrt{N}} a \psi^{\dagger}_i \psi_i
    \right],
\end{align}
where the summation over repeated indices $i = 1,\dots,N$ has been implicitly assumed.
All-in-all-out (AIAO) fluctuations are parametrized by an Ising order-parameter field $\phi$, which couples to the electronic quasiparticles as~\cite{savary14, boettcher17, ladovrechis21, moser24}
\begin{align}
    S_\phi = \int \dd{\tau} \dd[3]{\vec x} \left[
    \frac{1}{2} \phi ( r_1 -\nabla^2 ) \phi + \frac{g_1}{\sqrt{N}} \phi \psi^{\dagger}_i \gamma_{45} \psi_i
    \right],
\end{align}
where $r_1$ represents the tuning parameter for the AIAO transition and $g_1$ is the AIAO Yukawa coupling.
Nematic fluctuations are parametrized by a five-component tensor order-parameter field $\varphi_a$, which couples to the electronic quasiparticles as~\cite{janssen15, janssen16a, janssen17a}
\begin{align}
&S_{\varphi} = \int \dd{\tau} \dd[3]{\vec x} \Biggl[
\frac{1}{2} \sum_{a = 1}^5 \varphi_a ( r_2 - c \partial_\tau^2 - \nabla^2 ) \varphi_a 
\nonumber\\&
+ \frac{g_2}{\sqrt{N}} \sum_{a = 1}^5 \varphi_a \psi^{\dagger}_i \gamma_a \psi_i
%
%
+ \frac{\lambda}{\sqrt{N}} \sum_{a,b,c = 1}^5 \tr( \Lambda_a \Lambda_b \Lambda_c ) \varphi_a \varphi_b \varphi_c
\biggr]\,,
\end{align}
where $r_2$ represents the tuning parameter for the nematic transition, $g_2$ is the nematic Yukawa coupling, and $\lambda$ parametrizes the bosonic self-interaction.
$\Lambda_a$ denote the five real Gell-Mann matrices in three spatial dimensions.\footnote{%
We follow the notation of Ref.~\cite{moser24}. The five real Gell-Mann matrices, and therefore also the $\ell = 2$ real spherical harmonics used in Ref.~\cite{janssen15} correspond to the ones used here by relabelling indices $(1,2,3,4,5) \mapsto (4,3,2,1,5)$.}
The parameter $c$ is power-counting irrelevant and included for regularization purposes.
Quartic bosonic self-interactions are also power-counting irrelevant and are therefore set to zero from the beginning.

The fermion-boson model defined by the Euclidean action
\begin{align} \label{eq:fermion-boson-action}
S_\text{FB} = S_0 + S_a + S_\phi + S_\varphi
\end{align}
is quadratic in the Hubbard-Stratonovich fields $a$, $\phi$, and $\varphi$. The latter can therefore, in principle, be integrated out.
The original microscopic, purely fermionic action $S = S_0 + S_\text{int}$ presented in the main text can then be recovered from the above fermion-boson action in the limit of infinite order-parameter masses $r_{1,2} \to \infty$ and $g_{1,2}^2 \to \infty$, with fixed ratios
\begin{align} \label{eq:G1-G2}
G_1 = \frac{g_1^2}{r_1} \qquad \text{and} \qquad 
G_2 = \frac{g_2^2}{r_2},
\end{align}
and with $c = \lambda = 0$.

\subsection{Mean-field theory}
\label{subsec:mft}

In the large-$N$ limit and for small $G_1$ and $G_2$, the couplings flow to an infrared stable RG fixed point, characterizing an interacting semimetal phase. This phase, which realizes a three-dimensional non-Fermi liquid, is known as the Luttinger-Abrikosov-Beneslavskii (LAB) phase~\cite{abrikosov71, abrikosov74, moon13}.
Sizable four-fermion interactions beyond a finite threshold, however, drive instabilities of the LAB state, which can be understood within a mean-field analysis. To this end, we integrate out the fermion fields $\psi^\dagger_i$ and $\psi_i$, $i=1,\dots,N$, in Eq.~\eqref{eq:fermion-boson-action}, leading to an effective action for the Hubbard-Stratonovich fields. In the large-$N$ limit, the saddle point of this effective action dominates the partition function.
Determining the saddle-point solution ultimately reduces to minimizing the mean-field energy~\cite{janssen15, ray21, moser24},
\begin{align} \label{eq:mean-field-energy}
    E_\text{MF} (\phi, \varphi) & =
    \frac{r_1}{2} \phi^2 + \frac{r_2}{2} \varphi^2
    \nonumber \allowdisplaybreaks[1] \\ & \quad
    + N \int_0^\Lambda \frac{\dd[3]{\vec p}}{(2 \pi)^3} 
    \left[\varepsilon_{\phi, \varphi}^{+}(\vec p)
    + \varepsilon_{\phi, \varphi}^{-} (\vec p) \right]
\end{align}
with respect to the order-parameter fields $\phi$ and $\varphi$. Here, we used the fact that minimization of the mean-field energy implies the Coulomb field vanishes at the mean-field level, and assumed that the nematic order parameter takes the form $(\varphi_a) = (0, 0, 0, 0, \varphi)$. This form implies that the nematic phase is uniaxial, characterized by a single director with a residual discrete rotational symmetry remaining intact in the plane perpendicular to the director, and is believed to realize the smallest mean-field energy among the different possible nematic states~\cite{janssen15}.
$\Lambda$ is the ultraviolet cutoff.
The integrands $\varepsilon^{\pm}_{\phi,\varphi}(\vec p)$ denote the two lower-branch energy eigenvalues of the mean-field Hamiltonian
$
H_\text{MF} = \sum_{a = 1}^5 (1 + s_a \delta) d_a(\vec{p}) \gamma_a + \frac{g_1}{\sqrt{N}} \phi \gamma_{45} + \frac{g_2}{\sqrt{N}} \varphi \gamma_5
$.
They are given as
\begin{align}
    \varepsilon_{\phi, \varphi}^{\pm} (\vec{p}) & =
    -\vast[ (1 - \delta)^2 p^4 + 4 \delta \sum_{a=1}^3 d_a^2 (\vec{p}) + \left( \tfrac{g_1 \phi}{\sqrt{N}} \right)^2
    + \left( \tfrac{g_2 \varphi}{\sqrt{N}} \right)^2
    \nonumber \allowdisplaybreaks[1] \\ & 
    + 2 (1 - \delta) d_5 \tfrac{g_2 \varphi}{\sqrt{N}} 
    \pm 2 (1 + \delta) \tfrac{|g_1 \phi|}{\sqrt{N}} \sqrt{\sum_{a=1}^3 d_a^2 (\vec{p})}\vast]^{1/2}.
\end{align}
In Eq.~\eqref{eq:mean-field-energy}, the first two terms, proportional to $r_1$ and $r_2$, penalize finite order parameters, while the integral can lower the energy by causing partial or full band gap opening.

\begin{figure}[tb]
\centering
\includegraphics[width=\linewidth]{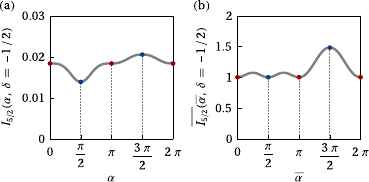}
\caption{(a) Coefficient $I_{5/2}$ of the nonanalytic term $\propto \chi^{5/2}$ in the mean-field energy, for fixed anisotropy parameter $\delta = -1/2$, as function of polar angle $\alpha$ in $(\phi, \varphi)$ field space. Here, $\alpha \in \{0, \pi\}$ ($\alpha \in \{\pi/2, 3\pi/2\}$) corresponds to the AIAO (nematic) sector.
$I_{5/2}$ is positive for all $\alpha \in [0,2\pi)$, justifying the truncation of the mean-field energy after the nonanalytic term.
(b) Rescaled coefficient $\overline{I_{5/2}}$ as function of rescaled polar angle $\overline \alpha$, for fixed $\delta = -1/2$, unveiling three global minima at $\overline\alpha = 0, \pi$ (AIAO sector) and $\overline\alpha = \frac{\pi}{2}$ (nematic sector).
This excludes the possibility of coexisting AIAO and nematic order.}
\label{fig:integral52}
\end{figure}

In order to expand the mean-field energy around small values of the order parameters, we introduce polar coordinates in field space as
$\phi = (\sqrt{N}/g_1) \chi \cos \alpha$
and
$\varphi = (\sqrt{N}/g_2) \chi \sin \alpha$,
with radial part $\chi > 0$ and angle part $\alpha \in [0, 2\pi)$.
A finite $\chi$ signals an instability of the Luttinger semimetal state, while the minimizing angle determines whether AIAO order ($\alpha \in \{0,\pi\}$), nematic order ($\alpha \in \{\pi/2, 3\pi/2\}$), or a coexistence of both orders is realized ($\alpha \notin \{0, \pi/2, \pi, 3\pi/2\}$).
Following Refs.~\cite{janssen15,moser24}, it is convenient to add two suitably written zeros to the mean-field energy, corresponding to the constant and quadratic terms in $E_\text{MF}$ with respect to $\chi$, namely
\begin{align}
    0 & = \int \frac{\dd{\Omega}}{(2 \pi)^3} 
    \vast[ \frac{1}{\tilde{X}^{3/4}} \int_0^{\frac{\tilde{X}^{1/4} \Lambda}{\chi^{1/2}}} \dd{x} \left( x^4 + \frac{c_\pm x^2}{2 \tilde{X}^{1/2}} + \frac{1}{2} - \frac{c_\pm^2}{8 \tilde{X}} \right) \chi^{5/2}
    \nonumber \allowdisplaybreaks[1] \\ & \quad
    - \left( \frac{\tilde{X}^{1/2} \Lambda^5}{5} + \frac{c_\pm \Lambda^3 \chi}{6 \tilde{X}^{1/2}} + \frac{\Lambda \chi^2}{2 \tilde{X}^{1/2}} - \frac{c_\pm^2 \Lambda \chi^2}{8 \tilde{X}^{3/2}} \right) \vast]\,,
\end{align}
where $c_\pm \coloneqq 2 (1 - \delta) \tilde{d}_5 \sin \alpha \pm 2 (1 + \delta) \abs{\cos \alpha} \sqrt{\sum_{a=1}^3 \tilde{d}_a^2}$ with $\tilde{d}_a \equiv \tilde d_a (\theta,\phi) \coloneqq d_a/p^2$ the real spherical harmonics, $\int \dd{\Omega}$ the integration over the solid angle in momentum space, and we abbreviated $\tilde{X} (\theta, \phi) \coloneqq (1-\delta)^2 + 12 \delta \sum_{i<j} \frac{p_i^2}{p^2} \frac{p_j^2}{p^2}$.
Adding those zeros allows us to express the mean-field energy for small $\chi \ll \Lambda^2$ as
\begin{align}  \label{eq:mean-field-energy-expanded}
    \frac{E_\text{MF} (\chi, \alpha)}{N \Lambda^5} & = \frac{1}{2} \left( \frac{r_1 \cos^2 \alpha}{g_1^2 \Lambda} + \frac{r_2 \sin^2 \alpha}{g_2^2 \Lambda} - I_2 (\alpha, \delta) \right) \left(\frac{\chi}{\Lambda^2}\right)^2
    \nonumber \allowdisplaybreaks[1] \\&\quad
    + I_{5/2} (\alpha, \delta) \left(\frac{\chi}{\Lambda^2}\right)^{5/2}
    + \mathcal{O} \left( \left(\frac{\chi}{\Lambda^2}\right)^{3} \right),
\end{align}
where we have subtracted an additive constant, which is irrelevant for the minimization problem, and introduced the dimensionless integrals
\begin{align}
    I_2 (\alpha, \delta) & \coloneqq \int \frac{\dd{\Omega}}{(2 \pi)^3} \left[ -\frac{c_{+}^2+c_{-}^2}{4\tilde{X}^{3/2}} +\frac{2}{\tilde{X}^{1/2}} \right]
    \allowdisplaybreaks[1] \\
    & = \frac{5 f_1 (\delta) - 3 (1+\delta) f_{2 \text{t}} (\delta)}{5 \pi^2} \cos^2 \alpha
    \nonumber \allowdisplaybreaks[1] \\& \quad
    + \frac{5 f_1 (\delta) - (1-\delta) f_{2 \text{e}} (\delta)}{5 \pi^2} \sin^2 \alpha\,,
    \allowdisplaybreaks[1] \\
    I_{5/2} (\alpha, \delta) & \coloneqq \int_0^\infty \dd{x} \int \frac{\dd{\Omega}}{(2 \pi)^3} \frac{1}{\tilde{X}^{3/4}} 
    \Bigg[ 2x^4 +\frac{1}{2}\frac{c_{+}+c_{-}}{\tilde{X}^{1/2}}x^2
    \nonumber \allowdisplaybreaks[1] \\& \quad
    -\frac{1}{8}\left(\frac{c_{+}}{\tilde{X}^{1/2}}\right)^2 -\frac{1}{8}\left(\frac{c_{-}}{\tilde{X}^{1/2}}\right)^2 +1
    \nonumber \allowdisplaybreaks[1] \\& 
    -x^2\sqrt{x^4+\frac{c_{+}}{\tilde{X}^{1/2}}x^2+1}
    -x^2\sqrt{x^4+\frac{c_{-}}{\tilde{X}^{1/2}}x^2+1} \Bigg]\,.
\end{align}
The integral $I_2 (\alpha, \delta)$ can be computed analytically and leads to anisotropy-dependent renormalizations of the tuning parameters $r_1$ and $r_2$.
The involved dimensionless functions $f_1$, $f_{2 \text{e}}$, and $f_{2 \text{t}}$ stem from solid-angle integration and are defined in \ref{app:solid-angle-integrals}.
Notably, the coefficient of the quadratic term in $\chi$ in Eq.~\eqref{eq:mean-field-energy-expanded} takes the form $r_1' \cos^2\alpha + r_2' \sin^2\alpha$ with dimensionless renormalized tuning parameters $r_{1,2}'$. Consequently, the coefficient is minimized for $\alpha = 0,\pi$ if $r_1' < r_2'$ and for $\alpha = \pi, 3\pi/2$ if $r_1' > r_2'$.
The nonanalytic term $I_{5/2} (\alpha, \delta) \chi^{5/2}$ arises from the presence of gapless fermionic degrees of freedom. For general $(\alpha, \delta)$, it has to be calculated numerically. Figure~\ref{fig:integral52}(a) shows $I_{5/2}$ as function of $\alpha$ for a representative fixed value of $\delta = -1/2$.
%
%
Importantly, $I_{5/2}(\alpha, \delta = -1/2) > 0$ for all values of $\alpha$.
This implies a continuous transition between the disordered Luttinger semimetal and the long-range-ordered phases, such that the higher-order terms $\mathcal O(\chi^3)$ in Eq.~\eqref{eq:mean-field-energy-expanded} can be neglected near the transition.
In order to determine the nature of the transition between the two ordered phases, it is useful to rescale the fields $\phi$ and $\varphi$ in a way that the coefficient of the nonanalytic term in the AIAO sector is the same as those in the nematic sector.
This can be achieved by introducing the rescaled polar angle $\overline\alpha$ using the field redefinitions
$\phi = [\sqrt{N}/(a_1 g_1)] \chi \cos \overline\alpha$
and
$\varphi = [\sqrt{N}/(a_2 g_2)] \chi \sin \overline\alpha$
with $a_1 \coloneqq [I_{5/2} (\alpha = 0)]^{2/5}$ and $a_2 \coloneqq [I_{5/2} (\alpha = \pi/2)]^{2/5}$.
Note that $\alpha = \overline\alpha$ for $\alpha \in \{0, \pi/2, \pi, 3\pi/2\}$.
On the level of the quadratic term $\propto \chi^2$ in the mean-field energy, the rescaling can be absorbed in a redefinition of the tuning parameters $r_1$ and $r_2$.
The rescaled coefficient of the nonanalytic term $\propto \chi^{5/2}$ now takes the form
\begin{align}
    \overline{I_{5/2}} ( \overline{\alpha},\delta) & \coloneqq
    \left( \frac{\cos^2 \overline{\alpha}}{a_1^2} + \frac{\sin^2 \overline{\alpha}}{a_2^2} \right)^{5/4} I_{5/2} ( \alpha(\overline\alpha), \delta)\,.
\end{align}
Figure~\ref{fig:integral52}(b) shows $\overline{I_{5/2}}$ as function of $\overline\alpha$ for a representative fixed value of $\delta = -1/2$.
The rescaled coefficient hosts three global minima, lying at angles $\overline{\alpha} = 0, \pi$ (AIAO sector) as well as $\overline{\alpha} = \pi/2$ (nematic sector).
This implies that a coexistence phase (corresponding to $\alpha \notin \{0, \pi/2, \pi, 3\pi/2\}$) is not realized for all values of $r_1'$ and $r_2'$. As a consequence, the mean-field theory predicts a direct discontinuous transition between the AIAO and nematic orders.

\begin{figure*}[t!]
\includegraphics[width=\linewidth]{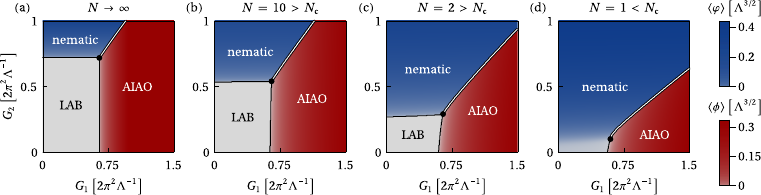}
\caption{%
(a)~Phase diagram of Luttinger model in the low-temperature limit as function of short-range couplings $G_1$ and $G_2$ for fixed representative values of the charge $e^2 = 3 \pi^2 \Lambda/2$ and the anisotropy parameter $\delta = -1/2$ from mean-field theory, which becomes exact in the limit $N \to \infty$. The Luttinger-Abrikosov-Beneslavskii (LAB) phase at small couplings realizes a three-dimensional non-Fermi liquid (gray). For sufficiently large $G_1$, a Weyl semimetal emerges, characterized by all-in-all-out (AIAO) antiferromagnetic order on the pyrochlore lattice. For sizable $G_2$, a nonmagnetic nematic topological insulator is stabilized.
Color scale indicates magnitudes of AIAO order parameter $\langle \phi \rangle \propto \langle \psi^\dagger \gamma_{45} \psi \rangle$ (red) and nematic order parameter $\langle \varphi \rangle \propto \langle \psi^\dagger \gamma_{5} \psi \rangle$ (blue). Single (double) lines indicate continuous (discontinuous) phase transitions.
(b)~Same as~(a), but for $N=10$ from RG analysis, qualitatively agreeing with the mean-field result.
(c)~Same as~(b), but for $N=2$. The LAB phase shrinks in favor of the nematic phase as the LAB fixed point and the quantum critical fixed point associated with the nematic instability approach each other.
(d)~Same as~(b), but for $N=1$, corresponding to the case relevant for the pyrochlore iridates $R_2$Ir$_2$O$_7$. The LAB phase is removed as a consequence of the fixed-point annihilation, leaving behind a continuous phase transition line between the AIAO antiferromagnetic Weyl semimetal and the nematic topological insulator for small $G_2$.
}
\label{fig:phasediagram}
\end{figure*}

This analytical result is consistent with the numerical minimization of the mean-field energy, Eq.~\eqref{eq:mean-field-energy-expanded}, with respect to the mean-field parameters $\chi$ and $\alpha$ for given $G_1 = g_1^2/r_1$, $G_2 = g_2^2/r_2$, and $\delta$.
This is illustrated in Fig.~\ref{fig:phasediagram}(a), which shows the resulting mean-field phase diagram as function of $G_1$ and $G_2$ for a fixed representative value of $\delta$. 
Beyond the LAB phase at small couplings, we identify two ordered phases:
For sufficiently large $G_1$, a Weyl semimetal phase emerges, characterized by the Ising order parameter $\langle \phi \rangle \propto \langle \psi^\dagger \gamma_{45} \psi \rangle$, which signals spontaneous time-reversal symmetry breaking. On the pyrochlore lattice, this phase exhibits all-in-all-out (AIAO) antiferromagnetic order~\cite{savary14, boettcher17, moser24}.
In contrast, for sizable $G_2$, we find a nonmagnetic phase that breaks lattice rotational symmetry. This phase is characterized by the nematic order parameter $\langle \varphi \rangle \propto \langle \psi^\dagger \gamma_5 \psi \rangle$, constructed from one of the five irreducible components of traceless symmetric tensors in three dimensions~\cite{janssen15}. It corresponds to a nematic topological insulator~\cite{herbut14, janssen17a}.
While the transitions from the LAB phase to the ordered phases are continuous, the direct transition between the two ordered phases is first order, consistent with the Landau-Ginzburg-Wilson paradigm.

\subsection{RG analysis}
\label{subsec:rg}

At finite $N$, fluctuation effects can become significant.
To account for these effects, we use an RG analysis in the dynamical bosonization framework~\cite{gies02, pawlowski07, floerchinger09}.
This approach applies a Hubbard-Stratonovich transformation after each RG step, effectively capturing potential nonperturbative effects from the RG-induced generation of four-fermions couplings while avoiding the unphysical singularities associated with their runaway flow.
The method has been successfully applied to Luttinger semimetals~\cite{janssen17a}, demonstrating the existence of a critical value $\Nc$ of quadratic band touching points at the Fermi level, consistent with earlier analyses~\cite{herbut14, janssen16a}. For $N \searrow \Nc$, the infrared stable fixed point corresponding to the LAB phase collides and annihilates with the quantum critical fixed point that governs the transition between the LAB phase and the nematic topological insulator.
All the different approaches predict a value of $\Nc$ around $2$, and therewith above the physical case $N=1$ relevant for the pyrochlore iridates.

Crucially, the dynamically bosonized RG approach allows us to compute the order parameters as functions of the microscopic couplings by integrating out the RG flow. This enables us to determine the nature of the phase transitions explicitly.

\subsubsection{$4-\epsilon$ expansion.}
\label{subsubsec:4-eps-expansion}
%
To begin with, note that the fermion-boson model defined in Eq.~\eqref{eq:fermion-boson-action} features a unique upper critical dimension. This is because the effective charge $e$, the AIAO Yukawa coupling $g_1$, the nematic Yukawa coupling $g_2$, and the nematic self-interaction $\lambda$ become simultaneously marginal in $d=4$ spatial dimensions. The presence of a unique upper critical dimension enables a controlled $4-\epsilon$ expansion.
To generalize our theory to noninteger spatial dimensions $2 < d < 4$, we maintain the general dimensional scaling of the couplings, but perform angular integrations directly in the physical dimension $d = 3$~\cite{vojta00a}. This approach is considered the most suitable for models of this type~\cite{janssen15}. By integrating out modes with momenta $q$ in the thin shell $\Lambda/b < q < \Lambda$, where $\Lambda$ is the ultraviolet cutoff, and arbitrary frequencies, the renormalized action takes the form
\begin{align}
    S_{\text{FB}}^{<} & = \int\limits_{-\infty}^\infty \frac{\dd{\omega}}{2 \pi} \int\limits_0^{\Lambda/b} \frac{\dd[d]{\vec q}}{(2 \pi)^d} \Bigg\{ 
    \sum_{i=1}^N \psi_i^\dagger \bigg[ 
    b^{\eta_1} \rmi \omega 
    \allowdisplaybreaks[1] \nonumber\\&\quad
    + b^{\eta_\psi} \sum_{a=1}^5 d_a(\vec{q}) \gamma_a + (\delta + \Delta \delta) \sum_{a=1}^5 s_a d_a(\vec{q}) \gamma_a \bigg] \psi_i
    \allowdisplaybreaks[1] \nonumber\\&\quad
    + \frac{1}{2} a b^{\eta_a} q^2 a
    + \frac{1}{2} \phi \left[ b^{\eta_\phi} q^2 + (r_1+\Delta r_1)\right] \phi
    \allowdisplaybreaks[1] \nonumber\\&\quad
    + \frac{1}{2} \sum_{a=1}^5 \varphi_a \left[ b^{\eta_\varphi} q^2 + (c + \Delta c) \omega^2 + (r_2+\Delta r_2)\right] \varphi_a
    \Bigg\} 
    \allowdisplaybreaks[2] \nonumber\\&\quad
    +\int\limits_{-\infty}^\infty \frac{\dd{\omega_1} \dd{\omega_2}}{(2 \pi)^2} \int\limits_0^{\Lambda/b} \frac{\dd[d]{\vec q_1} \dd[d]{\vec q_2}}{(2 \pi)^{2d}} \Bigg[
    \rmi \frac{e + \Delta e}{\sqrt{N}} a \sum_{i=1}^N \psi_i^\dagger \psi_i
    \allowdisplaybreaks[1] \nonumber\\&\quad
    + \frac{g_1 + \Delta g_1}{\sqrt{N}} \phi \sum_{i=1}^N \psi_i^\dagger \gamma_{45} \psi_i
    \allowdisplaybreaks[1] \nonumber\\&\quad
    + \frac{g_2 + \Delta g_2}{\sqrt{N}} \sum_{a=1}^5 \varphi_a \sum_{i=1}^N \psi_i^\dagger \gamma_a \psi_i
    \allowdisplaybreaks[1] \nonumber\\&\quad
    + \frac{\lambda + \Delta \lambda}{\sqrt{N}} \sum_{a,b,c = 1}^5 \tr( \Lambda_a \Lambda_b \Lambda_c ) \varphi_a \varphi_b \varphi_c
    \Bigg] .
\end{align}

\begin{figure}[bt!]
\centering
\includegraphics[width=\linewidth]{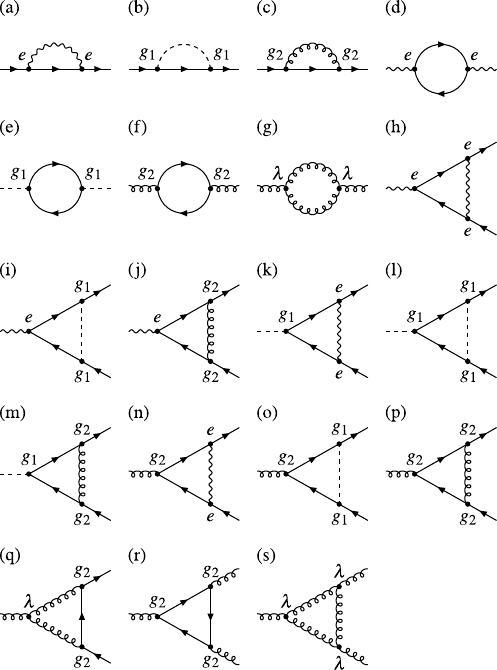}
\caption{Feynman diagrams at one-loop order contributing to
(a)--(c)~the fermion anomalous dimensions $\eta_1$, $\eta_\psi$ and the anisotropy parameter renormalization $\Delta\delta$,
(d)~the Coulomb anomalous dimension $\eta_a$,
(e)~the AIAO order-parameter anomalous dimension $\eta_\phi$ and the AIAO tuning parameter renormalization $\Delta r_1$,
(f,g)~the nematic order-parameter anomalous dimension $\eta_\varphi$, the parameter renormalization $\Delta c$, and the nematic tuning parameter renormalization $\Delta r_2$,
(h)--(j)~the charge renormalizations $\Delta e$,
(k)--(m)~the AIAO Yukawa vertex renormalizations $\Delta g_1$,
(n)--(p)~the nematic Yukawa vertex renormalizations $\Delta g_2$,
(q)--(s)~the self-interaction renormalizations $\Delta \lambda$.
The contributions to the flow of $e^2$ from (a) and (h), from (b) and (i), and from (c) and (j), respectively, cancel as a consequence of a Ward identity.}
\label{fig:diagrams-epsilon}
\end{figure}

We rescale all frequencies as $b^z \omega \mapsto \omega$, with dynamical exponent $z = 2 + \eta_1 - \eta_\psi$, and all momenta as $b \vec{q} \mapsto \vec{q}$.
Further, we renormalize the fields according to $b^{-(2+d+z-\eta_\psi)/2} \psi \mapsto \psi$, $b^{-(2+d+z-\eta_a)/2} a \mapsto a$, $b^{-(2+d+z-\eta_\phi)/2} \phi \mapsto \phi$, and $b^{-(2+d+z-\eta_\varphi)/2} \varphi_a \mapsto \varphi_a$.
The effective action then becomes of the same form as the original action, but with renormalized parameters $\delta$, $e$, $g_1$, $r_1$, $g_2$, $\lambda$, $c$, and $r_2$.
We take into account all one-loop self-energy and vertex diagrams depicted in Fig.~\ref{fig:diagrams-epsilon}.
%
%
The resulting flow equations read
\begin{align}
    \frac{\rmd\delta}{\rmd \ln b} & = - \eta_\psi \delta + \frac{2}{15 N} (1-\delta^2) [(1+\delta) f_{1 \text{t}} - (1-\delta) f_{1 \text{e}}] e^2
    \nonumber\\&\quad
    - \frac{2}{15 N} (1-\delta^2) [(1+\delta) f_{1 \text{t}} + (1-\delta) f_{1 \text{e}}] \frac{g_1^2}{(1+r_1)^3}
    \nonumber\\&\quad
    + \frac{2}{5 N} [(1+\delta) f_{1 \text{t}} - (1-\delta) f_{1 \text{e}}] \frac{g_2^2}{(1+r_2)^3},
\allowdisplaybreaks[1] \\
    \frac{\rmd e^2}{\rmd \ln b} & = \left(z+2-d- \eta_a + \frac{2\delta}{1-\delta^2}\frac{\rmd \delta}{\rmd \ln b}\right) e^2, 
\allowdisplaybreaks[1] \\
    \frac{\rmd g_1^2}{\rmd \ln b} & = \left(6-d-z-\eta_\phi-2\eta_\psi + \frac{2\delta}{1-\delta^2}\frac{\rmd \delta}{\rmd \ln b}\right) g_1^2
    \nonumber\\&\quad
    + \frac{2}{5 N} (1-\delta) (1-\delta^2) f_{2 \text{e}} g_1^2 e^2
    \nonumber\\&\quad
    - \frac{2}{5 N} (1-\delta) (1-\delta^2) f_{2 \text{e}} \frac{g_1^4}{1+r_1}
    \nonumber\\&\quad
    - \frac{2}{5 N} (1-\delta) f_{2 \text{e}} \frac{g_1^2 g_2^2}{1+r_2}, 
\allowdisplaybreaks[1] \\
    \frac{\rmd r_1}{\rmd \ln b} & = (2-\eta_\phi) r_1 - \frac{4}{5} (1-\delta) (1-\delta^2) f_{2 \text{e}} g_1^2,
\allowdisplaybreaks[1] \\ \label{eq:flow-g2}
    \frac{\rmd g_2^2}{\rmd \ln b} & = (6-d-z^{(0)}-\eta_\varphi^{(0)}-2\eta_\psi^{(0)}) g_2^2 + \frac{12}{5 N} \frac{g_2^4}{1+r_2}
    \nonumber\\&\quad
    + \frac{4}{5 N} (1-\delta^2) g_2^2 e^2 - \frac{4 c_1}{5 N} (1-\delta^2) \frac{g_1^2 g_2^2}{1+r_1},
\allowdisplaybreaks[1] \\ \label{eq:flow-lambda}
    \frac{\rmd \lambda}{\rmd \ln b} & = \frac{1}{2} (6-d-z^{(0)}-3\eta_\varphi^{(0)}) \lambda - \frac{27}{4 N} \frac{\lambda^3}{\sqrt{c} (1+r_2)^{5/2}}
    \nonumber\\&\quad
    - \frac{\sqrt{3}}{35} g_2^3,
\allowdisplaybreaks[1] \\ \label{eq:flow-c}
    \frac{\rmd c}{\rmd \ln b} & = (2-2z^{(0)}-\eta_\varphi^{(0)}) c + \frac{2}{5} g_2^2 + \frac{21}{4 N} \frac{\sqrt{c} \lambda^2}{(1+r_2)^{5/2}},
\allowdisplaybreaks[1] \\ \label{eq:flow-r2}
    \frac{\rmd r_2}{\rmd \ln b} & = (2-\eta_\varphi^{(0)}) r_2 - \frac{8}{5} g_2^2 - \frac{21}{N} \frac{\lambda^2}{\sqrt{c} (1+r_2)^{3/2}},
\end{align}
with the anomalous dimensions given by
\begin{align}
    \eta_\psi & = \frac{2}{15 N} (1-\delta^2) [(1+\delta) f_{1 \text{t}} + (1-\delta) f_{1 \text{e}}] e^2
    \nonumber \\ & \quad
    - \frac{2}{15 N} (1-\delta^2) [(1+\delta) f_{1 \text{t}} - (1-\delta) f_{1 \text{e}}] \frac{g_1^2}{(1+r_1)^3}
    \nonumber \\ & \quad
    + \frac{2}{5 N} [(1+\delta) f_{1 \text{t}} + (1-\delta) f_{1 \text{e}}] \frac{g_2^2}{(1+r_2)^3}, 
    \label{eq:eta-psi-AIAO} \allowdisplaybreaks[1] \\
    \eta_\psi^{(0)} & = \frac{4}{15 N} (1-\delta^2) e^2 + \frac{4}{5 N} \frac{g_2^2}{(1+r_2)^3}, 
    \label{eq:eta-psi-nematic} \allowdisplaybreaks[1] \\
    \eta_a & = e^2 f_{e^2}, 
    \allowdisplaybreaks[1]  \\
    \eta_\phi & = g_1^2 f_{g^2}, 
    \label{eq:eta-phi} \allowdisplaybreaks[1] \\
    \eta_\varphi^{(0)} & = \frac{44}{35} g_2^2 + \frac{21}{4 N} \frac{\lambda^2}{\sqrt{c} (1+r_2)^{5/2}}, 
    \label{eq:eta-varphi} \allowdisplaybreaks[1] \\
    \eta_1 & = 0,
\end{align}
and the dynamical exponent takes the form
\begin{align} \label{eq:dynamical-exponents}
    z &= 2 - \eta_\psi,
    & 
    z^{(0)} & = 2 - \eta_\psi^{(0)}.
\end{align}
In the above flow equations and anomalous dimensions, we have rescaled the couplings as $(e^2,g_1^2) \Lambda^{-\epsilon}/ (2 \pi^2) \mapsto (1 - \delta^2) (e^2,g_1^2)$, which turned out convenient to assess the properties of the AIAO fixed point~\cite{moser24}.
We have further introduced the dimensionless couplings $(g_2^2,\lambda^2) \Lambda^{-\epsilon}/ (2 \pi^2) \mapsto (g_2^2,\lambda^2)$, the dimensionless tuning parameters $(r_1,r_2) \Lambda^{-2} \mapsto (r_1,r_2)$, and the dimensionless parameter $c \Lambda^2 \mapsto c$.
The dimensionless functions $f_i(\delta)$ are defined in \ref{app:solid-angle-integrals}.
Note that the self-energy contributions to the flow of $e^2$, depicted in Figs.~\ref{fig:diagrams-epsilon}(a)-(c), cancel with the explicit charge vertex renormalizations, shown in Figs.~\ref{fig:diagrams-epsilon}(h)-(j), as a consequence of the Ward identity associated with the minimalistic gauge transformation $\psi \mapsto \rme^{\rmi e \lambda(\tau)}\psi$, $a \mapsto a - \partial_\tau \lambda(\tau)$~\cite{janssen17a,moser24}.
The parameter $c$ is an irrelevant parameter that is kept in order to regularize purely bosonic loops in the nematic sector, shown in Figs.~\ref{fig:diagrams-epsilon}(g) and \ref{fig:diagrams-epsilon}(s). 
Note that such regularization parameter is not required in the AIAO sector~\cite{moser24}, since a cubic bosonic self-interaction is forbidden by the Ising symmetry of the AIAO order parameter field $\phi$.
In the above, we have expanded the flow equations assuming small $c \ll 1$.
As will be shown in Sec.~\ref{subsubsec:fixed-point-structure}, this assumption is consistent with $c$ being small at all fixed points of interest, namely, the LAB fixed point, the nematic fixed point, and the AIAO fixed point.

Following Ref.~\cite{janssen17a}, we neglect the anomalous contribution $\propto \sum_{i,j,k=1}^d \partial_i T_{ij} \partial_k T_{kj}$ in the nematic sector, allowed whenever the spatial dimension $d$ agrees with $\dim T$, the dimension of the nematic tensor field $T_{ij} = \sum_{a = 1}^{(d-1)(d+2)/2} \varphi_a \Lambda_{a,ij}$.

For simplicity, the flow equations for the nematic Yukawa coupling $g_2^2$, the self-interaction coupling $\lambda$, the regularization parameter $c$, and the nematic tuning parameter $r_2$ have further been expanded to lowest order in the anisotropy parameter $\delta$ around the isotropic case $\delta=0$. This is sufficient to describe the nematic sector, see Sec.~\ref{subsubsec:fixed-point-structure}.
For consistency, \emph{before} any rescalings of coupling constants, also the involved anomalous dimensions $\eta_\psi$ and $\eta_\varphi$ and the dynamical exponent $z$ have to be expanded to their leading orders in $\delta$ in these cases, $\eta_\psi^{(0)} = \eta_\psi\big\vert_{\delta=0}$, $\eta_\varphi^{(0)} = \eta_\varphi\big\vert_{\delta=0}$, and $z^{(0)} = z\big\vert_{\delta=0}$.
Note, however, that anisotropy-dependent coupling reparametrizations, such as the above rescaling $(e^2,g_1^2) \mapsto (1 - \delta^2) (e^2,g_1^2)$, can still lead to crucial anisotropy dependencies.
On the level of the $4-\epsilon$ expansion, this happens explicitly through two vertex corrections in the flow equation of $g_2^2$ now involving a factor of $1-\delta^2$ and implicitly in the flow equations of $g_2^2$, $\lambda$, and $c$ through the reparametrized anomalous dimension $\eta_\psi^{(0)}$ involving a factor of $1-\delta^2$, see Eq.~\eqref{eq:eta-psi-nematic}.
In general, the theory space is spanned by 13 different parameters, namely
the anisotropy parameter $\delta$, 
the effective charge $e$, 
the AIAO Yukawa coupling $g_1$,
the AIAO tuning parameter $r_1$,
two nematic Yukawa couplings $g_{2, \text{t}}$ and $g_{2, \text{e}}$ (associated with the $\text{T}_{2\text{g}}$ and $\text{E}_\text{g}$ components of the nematic tensor order parameter),
three cubic nematic self-interactions $\lambda_{\text{t} \text{t} \text{t}}$, $\lambda_{\text{t} \text{t} \text{e}}$, and $\lambda_{\text{e} \text{e} \text{e}}$,
two regularization parameters $c_\text{t}$ and $c_\text{e}$, 
and two nematic tuning parameters $r_{2,\text{t}}$ and $r_{2,\text{e}}$.
Note that $\text{T}_{2\text{g}} \otimes \text{E}_\text{g} \otimes \text{E}_\text{g} = 2 \text{T}_{1\text{g}} \oplus 2 \text{T}_{2\text{g}}$~\cite{dresselhaus07}, which forbids terms like $\varphi_1 \varphi_4 \varphi_5$ in the action. This implies the absence of a fourth cubic self-interaction governed by a coupling $\lambda_{\text{t} \text{e} \text{e}}$.
Expanding the flow in the nematic sector around the isotropic limit allows us to restrict the parameter space to an eight-dimensional subspace spanned by $\delta$, $e$, $g_1$, $r_1$, $g_2$, $\lambda$, $c$, and $r_2$. The cost of this simplification is the introduction of an additional free parameter $c_1$ in the flow of the nematic Yukawa coupling $g_2$.
However, the contribution involving $c_1$ is merely a biquadratic term, $\propto g_1^2 g_2^2$. As such, we can already anticipate that its impact on the flow will likely be quantitative rather than qualitative.
Indeed, as will be shown in Sec.~\ref{subsubsec:fixed-point-structure}, both the location as well as the scaling exponents of the LAB fixed point, the nematic fixed point, and the AIAO fixed point remain entirely unaffected by the value of $c_1$.
We therefore expect that disregarding the distinction between the $\text{T}_{2\text{g}}$ and $\text{E}_\text{g}$ components of the nematic tensor order parameter will not affect the universal physics.
For the numerical integration of the flow equations presented in Sec.~\ref{subsubsec:phase-diagram}, we choose $c_1 \equiv 1/5$, which represents the weighted average of the $\text{T}_{2\text{g}}$ and $\text{E}_\text{g}$ contributions $c_\text{1t} = +1$ and $c_\text{1e} = -1$.

The flow equations presented here generalize those previously derived for the individual AIAO~\cite{moser24} and nematic~\cite{janssen17a} sectors, reducing to these in the respective limits.

\subsubsection{Dynamical bosonization scheme.}

\begin{figure}[tb]
\centering
\includegraphics[width=\linewidth]{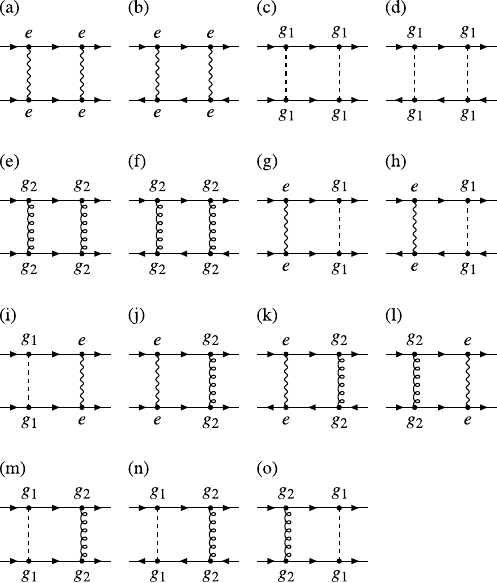}
\caption{Complete set of four-fermion box diagrams at the one-loop order. In the dynamical bosonization scheme, diagrams (a)-(f) and (m)-(o) contribute to the flow of the nematic Yukawa coupling $g_2$. In contrast, diagrams (g)-(i) and (j)-(l) do not contribute to the nematic channel when adding the individual contributions.}
\label{fig:diagrams-four-fermi}
\end{figure}

In three spatial dimensions, it is no longer sufficient to solely work with the flow equations arising from the $4-\epsilon$ expansion.
Additionally, we need to take into account four-fermion box diagrams, which generate further four-fermion interactions, not present at the microscopic scale. These are shown at the one-loop order in Fig.~\ref{fig:diagrams-four-fermi}.
Although these contributions can be safely neglected for small $\epsilon = 4-d \ll 1$, four-fermion contributions to the nematic channel turn out to be crucial in $2<d<4$ ~\cite{janssen17a}.
Neglecting these contributions would mask the fixed-point collision and annihilation of the LAB fixed point with the nematic fixed point when lowering the number of quadratic band touching points $N$ below a critical flavor number $N_\text{c}$.
Using dynamical bosonization in the nematic sector, RG-generated four-fermion interactions can be directly incorporated into the flow of the nematic Yukawa coupling $g_2^2$~\cite{gies02,pawlowski07,floerchinger09,janssen12,janssen17a}.
To illustrate this, consider the renormalized action
\begin{align}
    S_\text{FB}^< & = \int\limits_{\omega, \vec{q}} \frac{1}{2} \varphi_a (r_2+\Delta r_2) \varphi_a
    + \int\limits_{\omega_1, \vec{q_1}} \int\limits_{\omega_2, \vec{q_2}} \frac{g_2 + \Delta g_2}{\sqrt{N}} \varphi_a \psi_i^\dagger \gamma_a \psi_i
    \nonumber\\&\quad
    + \int\limits_{\omega_1, \vec{q_1}} \int\limits_{\omega_2, \vec{q_2}} \int\limits_{\omega_3, \vec{q_3}} \frac{\Delta G_2}{N} \left( \psi_i^\dagger \gamma_a \psi_i \right)^2
    + \dots,
\end{align}
where $\int\limits_{\omega, \vec{q}} \coloneqq \int_{-\infty}^\infty \frac{\dd{\omega}}{2 \pi} \int_0^{\Lambda/b} \frac{\dd[d]{\vec q}}{(2 \pi)^d}$ and the ellipsis denotes all other, tree level or RG generated, terms.
For notational simplicity, we assume summation convention over repeated indices in this section.
The renormalizations $\Delta r_2, \Delta g_2, \Delta G_2 =  \mathcal{O} (\ln b)$ contribute to the nematic order-parameter mass, the nematic Yukawa vertex, and the four-fermion coupling in the nematic channel, respectively.
Note that although the four-fermion interaction is absent in the ultraviolet, it gets generated under an RG transformation.
Shifting the nematic fields as $\varphi_a \mapsto \varphi_a + \frac{\Delta w}{\sqrt{N}} \psi_i^\dagger \gamma_a \psi_i$ results in the shifted renormalized action
\begin{align}
    S_\text{FB}^< & \mapsto \int\limits_{\omega, \vec{q}} \frac{1}{2} \varphi_a (r_2+\Delta r_2) \varphi_a
    \nonumber\\&\quad
    + \int\limits_{\omega_1, \vec{q_1}} \int\limits_{\omega_2, \vec{q_2}} \frac{g_2 + \Delta g_2 + r_2 \Delta w}{\sqrt{N}} \varphi_a \psi_i^\dagger \gamma_a \psi_i
    \nonumber\\&\quad
    + \int\limits_{\omega_1, \vec{q_1}} \int\limits_{\omega_2, \vec{q_2}} \int\limits_{\omega_3, \vec{q_3}} \frac{\Delta G_2 + g_2 \Delta w}{N} \left( \psi_i^\dagger \gamma_a \psi_i \right)^2
    + \dots,
\end{align}
assuming $\Delta w = \mathcal{O} (\ln b)$.
Choosing $\Delta w = - \Delta G_2 / g_2$ eliminates the four-fermion interaction at the cost of the additional Yukawa vertex renormalization $\propto - r_2 \Delta G_2 / g_2$.
In other words, the RG-generated four-fermion interaction can be incorporated through a Hubbard-Stratonovich transformation after each RG step, resulting in a modified flow equation for the Yukawa coupling.
Within the dynamical bosonization scheme, the flow equation of the nematic Yukawa coupling $g_2^2$ takes the form
\begin{align}
    \frac{\rmd g_2^2}{\rmd \ln b} & = (6-d-z^{(0)}-\eta_\varphi^{(0)}-2\eta_\psi^{(0)}) g_2^2 + \frac{12}{5 N} \frac{g_2^4}{1+r_2}
    \nonumber\\&\quad
    + \frac{4}{5 N} (1-\delta^2) g_2^2 e^2 - \frac{4 c_1}{5 N} (1-\delta^2) \frac{g_1^2 g_2^2}{1+r_1}
    \nonumber\\&\quad
    + \frac{1}{10 N} (1-\delta^2)^2 r_2 e^4 + \frac{1}{10 N} (1-\delta^2)^2 \frac{r_2 g_1^4}{(1+r_1)^2}
    \nonumber\\&\quad
    + \frac{13}{10 N} \frac{r_2 g_2^4}{(1+r_2)^2} - \frac{c_2}{N} (1-\delta^2) \frac{r_2 g_1^2 g_2^2}{(1+r_1)(1+r_2)},
\end{align}
where the first four summands stem from the $4-\epsilon$ expansion above and the last four terms are the contributions arising from the dynamical bosonization scheme.
Restricting to the eight-dimensional parameter space required us to introduce another free parameter $c_2$.
As with the previously encountered parameter $c_1$, the contribution corresponding to $c_2$ is also simply biquadratic, $\propto g_1^2 g_2^2$.
We show in Sec.~\ref{subsubsec:fixed-point-structure} that the location and the scaling exponents of the LAB fixed point, the nematic fixed point, and the AIAO fixed point are independent of both $c_1$ and $c_2$.
For the numerical integration of the flow equations, we choose, in analogy to the procedure for $c_1$, the weighted average $c_2 \equiv 2/5$ of the $\text{T}_{2\text{g}}$ and $\text{E}_\text{g}$ contributions $c_{2\text{t}} = 0$ and $c_{2\text{e}} = 1$.
The flow equations in the dynamical bosonization scheme reduce to the corresponding equations from Ref.~\cite{janssen17a} in the nematic sector when setting $g_1 = \delta = 0$.
As a further cross-check, we demonstrate in \ref{app:fermionic-flow} that the dynamically bosonized flow equations reduce, in the limit of infinite order-parameter masses, $r_1, r_2 \to \infty$, to those of a purely fermionic model, obtained without introducing order-parameter fields explicitly.

\subsubsection{Fixed-point structure.}
\label{subsubsec:fixed-point-structure}

\begin{table*}[t!]
    \caption{LAB fixed point: Fixed-point values and corresponding scaling exponents in $d = 3$ spatial dimensions from dynamical bosonization for different numbers of quadratic band touching points $N \geq N_\text{c} = 1.856$. Below the critical number $N_\text{c}$, the LAB fixed point has annihilated with the nematic fixed point and is no longer present in the physical space of real couplings.}
    \renewcommand{\arraystretch}{1.25}
    \begin{tabularx}{\textwidth}{Y | YYYYYYYY | YYYYYY}
    \hline\hline
        $N$ & $\delta_\star$ & $e_\star^2$ & $g_{1 \star}^2$ & $\alpha_{1 \star}$ & $g_{2 \star}^2$ & $\lambda_\star$ & $c_\star$ & $\alpha_{2 \star}$ & $2-z$ & $\eta_a$ & $\eta_\phi$ & $\eta_\varphi$ & $\omega_1$ & $\omega_2$ \\ \hline
        $N_\text{c}$ & 0.000 & 0.874 & 0.732 & 1.000 & 1.446 & $-0.040$ & 0.162 & 0.927 & 0.126 & 0.874 & 1.004 & 1.818 & 0.036 & 0.000 \\
        2 & 0.000 & 0.882 & 0.737 & 1.000 & 1.494 & $-0.040$ & 0.164 & 0.952 & 0.118 & 0.882 & 1.011 & 1.878 & 0.034 & 0.269 \\
        3 & 0.000 & 0.918 & 0.738 & 1.000 & 1.542 & $-0.040$ & 0.163 & 0.976 & 0.082 & 0.918 & 1.012 & 1.938 & 0.023 & 0.624  \\
        4 & 0.000 & 0.937 & 0.737 & 1.000 & 1.556 & $-0.040$ & 0.162 & 0.983 & 0.063 & 0.937 & 1.011 & 1.956 & 0.018 & 0.740 \\
        5 & 0.000 & 0.949 & 0.736 & 1.000 & 1.564 & $-0.040$ & 0.162 & 0.987 & 0.051 & 0.949 & 1.009 & 1.966 & 0.014 & 0.800 \\
        10 & 0.000 & 0.974 & 0.733 & 1.000 & 1.578 & $-0.040$ & 0.161 & 0.994 & 0.026 & 0.974 & 1.005 & 1.984 & 0.007 & 0.908 \\
        25 & 0.000 & 0.989 & 0.731 & 1.000 & 1.586 & $-0.040$ & 0.160 & 0.997 & 0.011 & 0.989 & 1.002 & 1.994 & 0.003 & 0.965 \\
        100 & 0.000 & 0.997 & 0.730 & 1.000 & 1.590 & $-0.040$ & 0.159 & 0.999 & 0.003 & 0.997 & 1.001 & 1.998 & 0.001 & 0.991 \\
        $\infty$ & 0 & 1 & $\frac{35}{48}$ & 1 & $\frac{35}{22}$ & $-\sqrt{\frac{105}{13310}}$ & $\frac{7}{44}$ & 1 & 0 & 1 & 1 & 2 & 0 & 1 \\
    \hline\hline
    \end{tabularx}
    \label{TableLABFixedPoint}
\end{table*}

\begin{table*}[t!]
    \caption{Nematic fixed point: Fixed-point values and corresponding critical exponents in $d = 3$ spatial dimensions from dynamical bosonization for different numbers of quadratic band touching points $N \geq N_\text{c} = 1.856$. Below the critical number $N_\text{c}$, the nematic fixed point has annihilated with the LAB fixed point and is no longer present in the physical space of real couplings.}
    \renewcommand{\arraystretch}{1.25}
    \begin{tabularx}{\textwidth}{Y | YYYYYYYY | YYYYY}
    \hline\hline
        $N$ & $\delta_\star$ & $e_\star^2$ & $g_{1 \star}^2$ & $\alpha_{1 \star}$ & $g_{2 \star}^2$ & $\lambda_\star$ & $c_\star$ & $\alpha_{2 \star}$ & $2-z$ & $\eta_a$ & $\eta_\phi$ & $\eta_\varphi$ & $1/\nu$ \\ \hline
        $N_\text{c}$ & 0.000 & 0.874 & 0.732 & 1.000 & 1.446 & $-0.040$ & 0.162 & 0.927 & 0.126 & 0.874 & 1.004 & 1.818 & 0.000 \\
        2 & 0.000 & 0.882 & 0.725 & 1.000 & 1.378 & $-0.039$ & 0.158 & 0.892 & 0.118 & 0.882 & 0.994 & 1.732 & 0.260 \\
        3 & 0.000 & 0.916 & 0.717 & 1.000 & 1.215 & $-0.038$ & 0.145 & 0.805 & 0.084 & 0.916 & 0.983 & 1.528 & 0.577 \\
        4 & 0.000 & 0.934 & 0.716 & 1.000 & 1.125 & $-0.037$ & 0.137 & 0.755 & 0.066 & 0.934 & 0.981 & 1.414 & 0.681 \\
        5 & 0.000 & 0.946 & 0.716 & 1.000 & 1.065 & $-0.037$ & 0.132 & 0.721 & 0.054 & 0.946 & 0.982 & 1.339 & 0.740 \\
        10 & 0.000 & 0.971 & 0.720 & 1.000 & 0.934 & $-0.036$ & 0.120 & 0.644 & 0.029 & 0.971 & 0.988 & 1.174 & 0.862 \\
        25 & 0.000 & 0.988 & 0.725 & 1.000 & 0.851 & $-0.035$ & 0.112 & 0.594 & 0.012 & 0.988 & 0.994 & 1.070 & 0.943 \\
        100 & 0.000 & 0.997 & 0.728 & 1.000 & 0.809 & $-0.035$ & 0.108 & 0.569 & 0.003 & 0.997 & 0.999 & 1.017 & 0.985 \\
        $\infty$ & 0 & 1 & $\frac{35}{48}$ & 1 & $\frac{35}{44}$ & $-\sqrt{\frac{105}{85184}}$ & $\frac{7}{66}$ & $\frac{14}{25}$ & 0 & 1 & 1 & 1 & 1 \\
    \hline\hline
    \end{tabularx}
    \label{TableNematicQCP}
\end{table*}

\begin{table*}[t!]
    \caption{AIAO fixed point: Fixed-point value and exact critical exponents in $d = 3$ spatial dimensions. The AIAO fixed point is present in the physical space of real couplings for any number of quadratic band touching points $N \in \mathbb{N}$.}
    \renewcommand{\arraystretch}{1.25}
    \begin{tabularx}{\textwidth}{Y | YYYYYYYY | YYYYY}
    \hline\hline
        $N$ & $\delta_\star$ & $e_\star^2$ & $g_{1 \star}^2$ & $\alpha_{1 \star}$ & $g_{2 \star}^2$ & $\lambda_\star$ & $c_\star$ & $\alpha_{2 \star}$ & $2-z$ & $\eta_a$ & $\eta_\phi$ & $\eta_\varphi$ & $1/\nu$ \\ \hline
        all & $-1$ & $\frac{9}{16}$ & $\frac{9}{16}$ & 0 & $\frac{35}{44}$ & $-\sqrt{\frac{105}{85184}}$ & $\frac{7}{66}$ & 1 & 0 & 1 & 1 & 1 & 1 \\
    \hline\hline
    \end{tabularx}
    \label{TableAIAOQCP}
\end{table*}

We now discuss the fixed-point structure of the model.
Based on the mean-field results, we expect that in the large-$N$ limit, the system exhibits a fully infrared stable fixed point corresponding to the disordered LAB phase~\cite{abrikosov71, abrikosov74, moon13}, along with two quantum critical fixed points associated with instabilities toward AIAO~\cite{savary14, boettcher17, moser24} and nematic~\cite{herbut14,janssen15,janssen17a} order, respectively.
For finite $N$, however, we explicitly show below that the LAB fixed point collides (annihilates) with the nematic fixed point at (below) a critical number of quadratic band touching points $N_\text{c} > 1$.
Moreover, we demonstrate that the AIAO fixed point does not take part in the collision and annihilation process, but remains in the physical space of real couplings for any number of quadratic band touching points $N \in \mathbb{N}$.

To observe the fixed-point annihilation at finite coupling strengths, it is convenient to reparametrize the AIAO Yukawa coupling $g_1^2 \mapsto g_1^2 / (- \delta)$ and the corresponding AIAO tuning parameter $r_1 \mapsto r_1 / (- \delta)$.
Additionally, we introduce the new tuning variables
\begin{align}
    \alpha_1 &\coloneqq \frac{r_1}{1 + r_1}, &
    \alpha_2 &\coloneqq \frac{r_2}{1 + r_2},
\end{align}
which map the AIAO and nematic tuning parameters $r_1$ and $r_2$ for $r_1, r_2 \in [0,\infty)$ onto the unit interval, $\alpha_1, \alpha_2 \in [0,1)$.
The fixed-point equations for the eight parameters $\delta, e^2, g_1^2, \alpha_1, g_2^2, \lambda, c, \alpha_2$ can be solved analytically in the limit $N \to \infty$.
Starting from the large-$N$ solutions, we use a Newton-Raphson method in order to determine the fixed-point values and the corresponding scaling or critical exponents numerically at finite $N$.
For sufficiently large $N$, we identify a unique interacting infrared stable fixed point, which we associate with the disordered LAB phase~\cite{abrikosov71, abrikosov74, moon13}.
The corresponding fixed-point values for different values of $N$, together with the associated scaling exponents, are given in Table~\ref{TableLABFixedPoint}.
In addition, as expected from the mean-field analysis, we further find two quantum critical fixed points associated with nematic~\cite{janssen15, janssen17a} and AIAO~\cite{savary14, boettcher17, moser24} instabilities. Their corresponding fixed-point values and critical exponents are given in Tables~\ref{TableNematicQCP} and \ref{TableAIAOQCP}, respectively.

As visible from Tables~\ref{TableLABFixedPoint} and \ref{TableNematicQCP}, the nematic fixed point and the LAB fixed point are located in the physical space of real couplings for $N \geq N_\text{c} = 1.8555070\dots$ (rounded to three decimal places for better readability in what follows).
At $N = N_\text{c}$, the LAB fixed point collides with the nematic fixed point and for even lower numbers of quadratic band touching points $N < N_\text{c}$, the two fixed points take complex values with finite imaginary parts.
This result is consistent with the earlier analyses~\cite{herbut14, janssen16a, janssen17a}. 
For the LAB fixed point, we show the dynamical exponent $z$, which is related to the fermion anomalous dimension as $\eta_\psi = 2 - z$, the photon anomalous dimension $\eta_a$, the anomalous dimension $\eta_\phi$ of the AIAO order parameter field, the anomalous dimension $\eta_\varphi$ of the nematic order parameter, as well as
the two least negative eigenvalues of the stability matrix at the LAB fixed point, $\omega_1$ and $\omega_2$, corresponding to leading and next-to-leading corrections to scaling. (We have labelled $\omega_1$ and $\omega_2$ such that $\omega_1 < \omega_2$ at large $N$, and smoothly connected to this limit for finite $N$.)
The scaling exponent $\omega_1$ reflects the fact that the anisotropy $\delta$ becomes an exactly marginal operator in the limit of infinitely many quadratic band touching points $N \rightarrow \infty$.
The scaling exponent $\omega_2 \rightarrow 0^+$ for $N \rightarrow N_\text{c}^+$ signals the collision of the LAB fixed point with the nematic fixed point.
For the nematic fixed point, we analogously show $z$, $\eta_a$, $\eta_\phi$, and $\eta_\varphi$, as well as the inverse correlation-length exponent $1/\nu$, corresponding to the unique positive eigenvalue of the stability matrix.
We note that the AIAO order parameter is non-critical at the nematic quantum critical point, indicated by $\alpha_{1\star} = 1$. The corresponding anomalous dimension $\eta_\phi$ at the nematic fixed point determines the scaling of the AIAO correlation function only after subtracting the momentum-independent contribution.
The fact that $\alpha_{2\star} \in (0,1)$ at the nematic fixed point implies that the corresponding dimensionful mass parameter $\Lambda^2 \alpha_{2\star}/(1-\alpha_{2\star})$ vanishes in the infrared limit, consistent with the emergent scale invariance at criticality.
The numerical values of $z$, $\eta_a$, $\eta_\varphi$, $\omega_2$, and $1/\nu$ in Tables~\ref{TableLABFixedPoint} and \ref{TableNematicQCP} precisely agree with the corresponding values given in Ref.~\cite{janssen17a}. This is a consequence of the fact that AIAO fluctuations are suppressed by the infinite mass of the AIAO order-parameter field $\phi$ at the LAB and nematic fixed points, as indicated by the corresponding $\alpha_{1\star} = 1$.

As shown in Table~\ref{TableAIAOQCP}, the AIAO fixed point exists for any number of quadratic band touching points $N \in \mathbb{N}$ and, therefore, is not involved in any fixed-point annihilations.
The fixed point is characterized by maximal anisotropy $\delta_\star = -1$, which has the consequence that the location of the fixed point  and the corresponding critical exponents can be computed analytically.
In fact, as argued in \ref{app:higher-loop-corrections}, the AIAO critical exponents are expected to be exact.
Our results for $z$, $\eta_a$, $\eta_\phi$, and $1/\nu$ also precisely agree with the earlier calculations~\cite{savary14,moser24}.

We have explicitly verified that locations and exponents of all three fixed points are independent from the parameters $c_1$ and $c_2$ introduced in Sec.~\ref{subsubsec:4-eps-expansion}.
This implies that also our result for the critical number $N_\text{c}$ at which the fixed-point collision takes place does depend on neither $c_1$ nor $c_2$.
This justifies a posteriori the earlier introduced restriction to an eight-dimensional theory space.

To summarize the fixed-point analysis, the flow equations in the dynamical bosonization scheme exhibit, at large $N$, an infrared stable fixed point corresponding to the disordered LAB phase, along with two quantum critical fixed points associated with instabilities toward the AIAO antiferromagnetic Weyl semimetal and the nematic topological insulator, respectively. As $N$ is lowered toward a critical value $N_\mathrm{c} = 1.856$, the stable LAB fixed point and the quantum critical nematic fixed point approach each other and ultimately annihilate for $N < N_\mathrm{c}$.
We now turn to the implications of this fixed-point annihilation for the phase diagram.

\subsubsection{Phase diagram.}
\label{subsubsec:phase-diagram}

To obtain the zero-temperature phase diagram, Figs.~\ref{fig:phasediagram}(b)--(d), we numerically integrate out the flow equations for the eight parameters $\delta, e^2, G_1, \alpha_1, G_2, \lambda, c, \alpha_2$  in $d = 3$ spatial dimensions, using varying initial effective interaction strengths in the AIAO (nematic) channel $G_1$ ($G_2$), and fixed 
$\left( \delta, 2e^2/(3\pi^2 \Lambda), \alpha_1, \lambda/\sqrt{2\pi^2\Lambda}, c \Lambda^2, \alpha_2 \right)_{\text{UV}}
=  \left( -\frac{1}{2}, 1, \frac{1}{10}, -\sqrt{\frac{105}{85184}}, \frac{7}{66}, \frac{14}{25} \right)$
at the ultraviolet scale.
Here, $G_1$ and $G_2$ correspond to the four-fermion couplings in the original four-fermion action in Eq.~\eqref{eq:LagrangianInt}, which are related to the Yukawa couplings $g_1^2$ and $g_2^2$ as
\begin{align}
G_1 = \frac{g_1^2}{r_1} \qquad \text{and} \qquad 
G_2 = \frac{g_2^2}{r_2},
\end{align}
cf.~Eq.~\eqref{eq:G1-G2}.
In order to allow for an easier comparison with the mean-field results and other studies, we have chosen to work with dimensionful quantities $e^2$, $G_1$, $G_2$, $\lambda$, and $c$ in this subsection.
The initial values of the effective charge $e^2$, the cubic nematic self-interaction $\lambda$, the regularization parameter $c$, and the nematic mass parameter $\alpha_2$ have been chosen according to the location of the nematic fixed point in the limit of infinitely many quadratic band touching points $N \rightarrow \infty$.
The factors of $2/(3\pi^2)$ and $1/\sqrt{2\pi^2}$ in the definition of the starting values for $e^2$ and $\lambda$ arise from the coupling rescaling for $\delta = -1/2$, as given after Eq.~\eqref{eq:dynamical-exponents}.
While the location of the phase boundaries depends on the choice of the initial values, our universal results, such as the occurrence of a continuous order-to-order transition is independent of this choice.
Depending on the chosen initial conditions, for $N > N_\text{c}$, we find a flow towards either of three different regimes:
For small $G_1$ and $G_2$, the parameters flow towards the LAB fixed-point values. In particular, this implies that $\alpha_1$ and $\alpha_2$ remain positive at all RG scales. This regime is associated with the LAB phase.
For positive $G_1$ beyond a certain finite threshold, we find that $\alpha_1$ becomes negative at some RG scale. Negative $\alpha_1$ corresponds to a negative curvature of the effective potential in the AIAO sector. Consequently, we associate the flow of $\alpha_1$ toward negative values with the onset of AIAO order.
For positive $G_2$ beyond a certain finite threshold, on the other hand, we find that $\alpha_2$ becomes negative at some RG scale, which we associate with the onset of nematic order.
In order to estimate the values of the order parameters in the two different ordered phases, we assume that the flow \emph{away} from the quantum critical fixed point is dominated by simple dimensional scaling. This assumption only modifies nonuniversal properties of the observables; in particular, it does neither change the order of the transition nor the values of the corresponding critical exponents~\cite{rosa01, janssen12, dupuis21}.
Under this assumption, the values of the order parameters can be computed from the RG time $t_\text{IR} = \ln b_\text{IR}$, at which the associated parameters $\alpha_1$ and $\alpha_2$, respectively, turn negative.
In particular, in the AIAO phase, the order parameter $\langle \phi \rangle$ is related to $t_\text{IR}$ via $\langle \phi \rangle \propto \exp(- \frac{d + z - 2 + \eta_\phi}{2} t_\text{IR}) = \exp(- 2 t_\text{IR})$, since $z = 2$ and $\eta_\phi = 1$ at the AIAO quantum critical fixed point for all numbers of quadratic band touching points $N \in \mathbb{N}$ and in $d =3$ spatial dimensions.
In the nematic phase, the corresponding order parameter $\langle \varphi \rangle$
can be computed analogously; however, one has to be careful which dynamical exponent and anomalous dimension to incorporate:
For $N = 10$ quadratic band touching points, we identify $\langle \varphi \rangle \propto \exp(- \frac{d + z - 2 + \eta_\varphi}{2} t_\text{IR}) = \exp(- 2.072 t_\text{IR})$, with $z = 1.971$ and $\eta_\varphi = 1.174$ at the nematic fixed point, see Table~\ref{TableNematicQCP}.
Likewise, for $N = 2$ quadratic band touching points, we employ again $\langle \varphi \rangle \propto \exp(- \frac{d + z - 2 + \eta_\varphi}{2} t_\text{IR}) = \exp(- 2.307 t_\text{IR})$, since $z = 1.882$ and $\eta_\varphi = 1.732$ at the nematic fixed point, see Table~\ref{TableNematicQCP}.
However, for the case $N = 1$, and therewith after the fixed-point annihilation, we identify $\langle \varphi \rangle \propto \exp(- \frac{d + z - 2 + \eta_\phi}{2} t_\text{IR}) = \exp(- 2 t_\text{IR})$, with $z = 2$ and $\eta_\phi = 1$ from the annihilation-surviving AIAO fixed point, which governs the transition into the nematic phase.

The resulting quantum phase diagram for a representative value of $N=10$ is shown in Fig.~\ref{fig:phasediagram}(b) and qualitatively aligns with the mean-field result in Fig.~\ref{fig:phasediagram}(a).
As $N$ decreases, the LAB fixed point and the quantum critical fixed point associated with the nematic instability approach each other.
This implies that the LAB phase shrinks in the phase diagram in favor of the nematic phase, as exemplified for $N=2$ in Fig.~\ref{fig:phasediagram}(c).
Upon further decreasing $N$, the two fixed points eventually collide at a critical value $\Nc = 1.856$, in agreement with Ref.~\cite{janssen17a}.
The fixed-point annihilation for $N < \Nc$ alters the flow topology by removing the LAB phase.
We emphasize that no additional fixed points are involved in the collision; in particular, the quantum critical fixed point associated with the AIAO instability remains within the physical space of real couplings for all $N \geq 1$.
The resulting quantum phase diagram for the physical case of $N=1$, relevant for the pyrochlore iridates, is shown in Fig.~\ref{fig:phasediagram}(d).
Importantly, there is now a direct transition between the AIAO Weyl semimetal and the nematic topological insulator upon tuning a single parameter, $G_1$.
Although the two ordered states break different symmetries, and the Landau-Ginzburg-Wilson paradigm would predict a discontinuous transition, the presence of the quantum critical fixed point associated with the AIAO instability leads to a continuous transition without fine tuning, provided $G_2$ is not too large.

\begin{figure}[tb!]
\includegraphics[width=\linewidth]{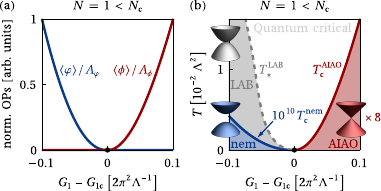}
\caption{%
(a)~Normalized order parameters (OPs) $\langle \varphi \rangle/A_\varphi$ and $\langle \phi \rangle/A_\phi$ for nematic (blue) and AIAO (red) order, respectively, for $N=1$ as functions of the ultraviolet value of the short-range coupling $G_1$ from RG analysis for representative fixed ultraviolet values of the short-range coupling $G_2 = 0$, charge $e^2 = 3 \pi^2 \Lambda/2$, and anisotropy parameter $\delta = - 1/2$.
There is a direct continuous transition at $G_\text{1c} = 0.565 \cdot 2 \pi^2/ \Lambda$ between the nematic topological insulator for $G_1 < G_\text{1c}$ and the AIAO Weyl semimetal for $G_1 > G_\text{1c}$.
The normalization factors were chosen such that $\langle \phi \rangle = A_\phi$ and $\langle \varphi \rangle = A_\varphi$ for $G_1 - G_\text{1c} = \pm 0.1 \cdot 2 \pi^2/ \Lambda$, respectively.
The order parameters scale as $\langle \phi \rangle \propto (G_1 - G_\text{1c})^{\beta_\phi}$ and $\langle \varphi \rangle \propto (G_\text{1c} - G_1)^{\beta_\varphi}$ for $G_1$ approaching $G_\text{1c}$ from above and below, respectively, with the same exponent $\beta_\phi = \beta_\varphi = 2$, but with significantly different prefactors.
(b)~Finite-temperature phase diagram for $N = 1$ as function of $G_1$ from RG analysis, with ultraviolet values of $G_2$, $e^2$, and $\delta$ as in (a).
Solid (dashed) lines denote phase transitions (crossovers).
The quantum critical point between the nematic topological insulator (blue) and the AIAO Weyl semimetal (red) is marked as black dot.
The slow RG flow resulting from fixed-point annihilation causes a striking anisotropy between the two sides of the transitions, with the critical temperature significantly suppressed on the nematic side.
This suppression of the nematic order leaves an intermediate region at finite temperature dominated by LAB-like behavior (gray).
}
\label{fig:opsandfinitetemp}
\end{figure}

To illustrate this fact, we show in Fig.~\ref{fig:opsandfinitetemp}(a) the nematic order parameter $\langle \varphi \rangle \propto \langle \psi^\dagger \gamma_5 \psi \rangle$ and the AIAO antiferromagnetic order parameter $\langle \phi \rangle \propto \langle \psi^\dagger \gamma_{45} \psi \rangle$ for $N=1$ as functions of the ultraviolet value of $G_1$ while keeping the ultraviolet values of $G_2$, $e^2$, and $\delta$ fixed. 
The two order parameters vanish continuously upon approaching the quantum critical point at $G_\text{1c}$ from above and below, respectively.
The same behavior is observed for other values of $G_2$, $e^2$, and $\delta$, provided $G_2$ is not too large. There is a direct continuous order-to-order quantum phase transition without an intermediate mixed phase and without additional fine tuning.

\subsubsection{Asymmetry in energy scales.}

\begin{figure*}[t!]
\includegraphics[width=\linewidth]{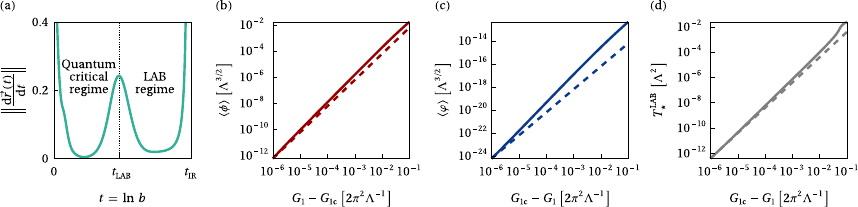}
\caption{%
(a)~Flow speed $\norm{\frac{\rmd \vec{r} (t)}{\rmd t}}$ as a function of RG time $t \in [0, t_\text{IR}]$ for $N=1$ and initial values of the flow parameters on the nematic side of the transition, 
$\vec r(t = 0) = 
( -\frac{1}{2}, 1, \frac{G_{1\text{c}}}{3\pi^2/(2\Lambda)} - \frac{4}{3} \cdot 10^{-5}, \frac{1}{10}, 0, -\sqrt{\frac{105}{85184}}, \frac{7}{66}, \frac{14}{25} )$, with $G_{1\text{c}} = 0.752\ldots \cdot  \frac{3\pi^2}{2 \Lambda}$.
The speed displays two regimes of slow flow, indicated by the two minima of $\norm{\frac{\rmd \vec{r} (t)}{\rmd t}}$. The first (second) regime arises from the vicinity to the AIAO (complex LAB) fixed point. The local maximum at intermediate $t_\text{LAB}$ corresponds to the crossover between the two regimes.
(b)~AIAO order parameter $\langle \phi \rangle$ as a function of $G_1 - G_{1\text{c}}$ in double logarithmic plot on the AIAO side of the transition for $N=1$.
All other initial flow parameters have been chosen as in (a).
Dashed line indicates the theoretically expected slope $\nu z =2$.
(c)~Same as (b), but for the nematic order parameter $\langle \varphi \rangle$ as a function of $G_{1\text{c}} - G_1$ on the nematic side of the transition. The exceptionally small value of the nematic order parameter is a result of the complex LAB fixed point still being reasonable close to the physical space of real couplings at $N \lesssim N_\text{c} = 1.856$, resulting in a slow flow and a sizable intermediate-temperature LAB-like regime.
(d)~Same as (c), but for the crossover temperature $T_\star^\text{LAB}$ between the nonuniversal high-temperature regime and the intermediate-temperature LAB-like regime on the nematic side of the transition.
}
\label{fig:couplingspeedandfits}
\end{figure*}

A characteristic feature of the continuous order-to-order transition from fixed-point annihilation is a significant anisotropy in the effective energy scales on the two sides of the transition.
In general, energy scales depend on microscopic parameters and therefore cannot be quantitatively predicted within our continuum model. However, at equal distances from criticality, we qualitatively find that the nematic order parameter for $G_1 < G_\text{1c}$ is typically several order of magnitude smaller than the AIAO order parameter for $G_1 > G_\text{1c}$.
This phenomenon arises as a direct consequence of the fixed-point annihilation occurring at $\Nc = 1.856$, which is only slightly above the physical value of $N=1$.
For $N$ just below $\Nc$, the complex LAB fixed point has a small imaginary part.
As a result, the RG flow near the location in parameter space where the fixed-point annihilation has occurred becomes logarithmically slow. This slow flow leads to an exponential suppression of the effective energy scale associated with the nematic order, a phenomenon known as ``walking behavior'' in the field-theory literature~\cite{kaplan09, gorbenko18a}. Walking behavior is observed in fixed-point annihilation scenarios across various systems, including quantum impurity models~\cite{nahum22,hu22,weber23,weber24}, Potts models~\cite{gorbenko18b, ma19, jacobsen24, tang24}, Abelian Higgs models~\cite{halperin74,ihrig19}, Wess-Zumino-Witten models~\cite{ma20, nahum20, hawashin25}, QED$_3$~\cite{braun14, janssen16b, herbut16a, gukov17}, and QCD$_4$~\cite{gies06,braun10,braun11,kusafuka11,kuipers19}.

To illustrate this point, we again assume that the flow away from the quantum critical fixed point is dominated by simple dimensional scaling. Up to prefactors of order one, we can then estimate the critical temperature on the AIAO side of the transition as $T_\text{c}^\text{AIAO} \propto \exp(- z t_\text{IR}) = \exp(- 2 t_\text{IR})$, where $t_\text{IR} = \ln b_\text{IR}$ denotes the RG time at which the parameter $\alpha_1$ turns negative.
Similarly, on the nematic side of the transition, we can estimate the nematic critical temperature as $T_\text{c}^\text{nem} \propto \exp(- 2 t_\text{IR})$, with $t_\text{IR}$ denoting the RG time at which the parameter $\alpha_2$ turns negative.
For $N = 1$, the complex LAB fixed point is still reasonably close to the physical space of real couplings.
This results in a regime with logarithmically slow flow and LAB-like behavior in a sizable finite-temperature regime~\cite{herbut14,janssen17a}.
The corresponding crossover temperature $T_\star^\text{LAB}$ between the nonuniversal high-temperature regime and the LAB-like regime at intermediate temperatures can be estimated via $T_\star^\text{LAB} \propto \exp(- 2 t_\text{LAB})$, where $t_\text{LAB}$ corresponds to the RG time at which the flow enters the slow-flow regime associated with the complex LAB fixed point.
A natural way to estimate $t_\text{LAB}$ is to analyze the dimensionless vector of flow parameters $\vec{r}(t) \coloneqq ( \delta, e^2/[2\pi^2(1-\delta^2) \Lambda], G_1 \Lambda /[2\pi^2(1-\delta^2)], \alpha_1, G_2 \Lambda/(2\pi^2), \lambda/\sqrt{2\pi^2\Lambda}, c \Lambda^2, \alpha_2 )(t)$ as function of RG time $t$.
Figure~\ref{fig:couplingspeedandfits}(a) shows the corresponding flow speed $\norm{\frac{\rmd \vec{r} (t)}{\rmd t}}$ as function of $t \in [0, t_\text{IR}]$.
We observe two regimes in which the flow becomes slow. Comparing the flow parameters $\vec{r}(t)$ with the values of the different fixed points, we identify the first (second) regime of slow flow to arise from the vicinity to the AIAO (complex LAB) fixed point.
The first minimum in $\norm{\frac{\rmd \vec{r} (t)}{\rmd t}}$ thus corresponds to the quantum critical regime, while the second minimum corresponds to the LAB regime.
The local maximum in $\norm{\frac{\rmd \vec{r} (t)}{\rmd t}}$ in between the two minima at intermediate time $t_\text{LAB}$ thus corresponds to the crossover between the two regimes.

In sum, on the nematic side of the transition, we find a sizable intermediate-temperature regime between the critical temperature $T_\mathrm{c}^\text{nem}$ and the crossover temperature $T_\star^\text{LAB}$.
On the AIAO side, there is only a single minimum in the flow speed $\norm{\frac{\rmd \vec{r} (t)}{\rmd t}}$, associated with the quantum critical regime, indicating the absence of an additional intermediate-temperature regime.
The resulting finite-temperature phase diagram is shown in Fig.~\ref{fig:opsandfinitetemp}(b), and illustrates the advertised asymmetry in the effective energy scales on the two different sides of the continuous order-to-order transition.
On the nematic side of the transition, the critical temperature is significantly suppressed by several orders of magnitude in comparison with the AIAO side. 
This suppression of the nematic order creates a broad intermediate regime at finite temperature, separated from the nonuniversal high-energy regime by a crossover scale $T_\star^\text{LAB}$. In this regime, the RG flow is ``stuck'' in the vicinity of the, now imaginary, LAB fixed point~\cite{herbut14,janssen17a}, and thermodynamic and transport observables exhibit power laws with nontrivial exponents, characteristic of a non-Fermi liquid~\cite{moon13}.
While the asymmetry in the effective energy scales on either side of the transition decreases with increasing $G_2$, we find it to remain significant for all parameter sets that we have studied.
This indicates that it can serve as a distinctive signature for experimentally identifying the proposed mechanism for continuous order-to-order transitions from fixed-point annihilation.

In the present example, the proximity of the physical value $N = 1$ to the critical value $\Nc = 1.856$ results in an asymmetry 
large enough to potentially hinder direct experimental verification. However, since the asymmetry diminishes with increasing distance from the quantum critical point, finite nematic order may still be observable, provided the system is sufficiently removed from the critical region, leading to indirect evidence of the proposed continuous order-to-order transition.
Another possible indirect signature of the mechanism is the appearance of ``drifting'' exponents in the pseudo-critical regime above the nematic phase, resulting from the fixed-point annihilation~\cite{herbut14, janssen17a}.
%

\subsubsection{Universality class.}

The quantum critical behavior of the continuous order-to-order transition is governed by the RG flow near the fixed point associated with the AIAO instability. 
In the present model, this fixed point is situated at extreme cubic anisotropy~\cite{savary14, moser24}, allowing for the exact computation of all critical exponents, see \ref{app:higher-loop-corrections}.
We find that the anomalous dimensions for both the AIAO and nematic orders, respectively are $\eta_\phi = \eta_\varphi = 1$, the correlation-length exponents $\nu_\phi = \nu_\varphi = 1$, and the dynamical exponent $z=2$.
Assuming hyperscaling holds, we find that the two order parameters scale as $\langle \phi \rangle \propto (G_1 - G_\text{1c})^{\beta_\phi}$ and $\langle \varphi \rangle \propto (G_\text{1c} - G_1)^{\beta_\varphi}$ as $G_1$ approaches $G_\text{1c}$ from above and below, respectively, with the same exponent $\beta_\phi = \beta_\varphi = 2$.

This theoretical expectation can be explicitly confirmed within our numerical analysis. To this end, we replot the data from Fig.~\ref{fig:opsandfinitetemp} using a double-logarithmic scale.
Figure~\ref{fig:couplingspeedandfits}(b) displays the AIAO order parameter as a function of the tuning parameter $G_1 - G_{1\text{c}}$ on the AIAO side of the transition.
Near the critical point, the values obtained from numerical integration (solid line) approach the expected behavior with slope $\beta_\phi = 2$ (dashed line), as expected from the fixed-point analysis.
Similarly, Fig.~\ref{fig:couplingspeedandfits}(c) displays the nematic order parameter as a function of $G_{1\text{c}} - G_1$ on the nematic side of the transition. While corrections to scaling appear to be more significant in this case, the leading behavior is still consistent with the expected slope $\beta_\varphi = 2$.
We attribute the sizable scaling corrections to the slow flow arising from the fixed-point annihilation and the marginal anisotropy parameter $\delta$, which leads to quasiuniversal behavior over several length scales~\cite{moser24}.
Lastly, Fig.~\ref{fig:couplingspeedandfits}(d) depicts the crossover temperature $T_\star^\text{LAB}$ as a function of $G_{1\text{c}} - G_1$ on the nematic side of the transition. Again, the numerical result (solid line) in the vicinity of the critical point is consistent with the expectation from the fixed-point analysis (dashed line), namely $T_\star^\text{LAB} \propto (G_{1\text{c}} - G_1)^{\nu z}$ with $\nu z = 2$.

Let us end this section with two general comments on the universality class of an order-to-order quantum phase transition caused by the proposed fixed-point annihilation mechanism.
First, it is of course no coincidence that the two correlation-length exponents $\nu_\phi$ and $\nu_\varphi$ associated with the AIAO and nematic correlation functions align in our theory, $\nu = \nu_\phi = \nu_\varphi$.
This has to hold for any order-to-order quantum phase transition arising from fixed-point annihilation as a consequence of the presence of a single divergent length scale at the annihilation-surviving quantum critical fixed point.
Second, all other pairs of critical exponents associated with the two different order parameters may in general take on different values. The fact that this does not happen in the present case can be understood as a consequence of the maximal anisotropy $\delta_\star = -1$ at the AIAO fixed point, leading to $\eta_\phi = \eta_\varphi = 1$, which, together with the hyperscaling assumption, implies that all pairs of exponents agree.

\section{Kagome quantum magnets}
\label{sec:kagome}

\begin{figure*}[t!]
\includegraphics[width=\linewidth]{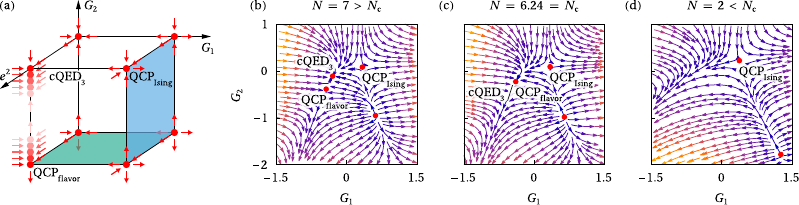}
\caption{%
(a) Schematic fixed-point structure of the QED$_3$-Gross-Neveu model in the space spanned by the couplings $e^2$, $G_1$, and $G_2$.
Arrows indicate flow towards infrared.
For sufficiently large $N$, the infrared attractive plane $e^2 = e^2_\star$ hosts a fully infrared stable fixed point (cQED$_3$), associated with the conformal phase of QED$_3$, representing the U(1) Dirac spin liquid state in the kagome Heisenberg spin-1/2 model.
In addition, the plane features two quantum critical fixed points, associated with instabilities towards $\mathbb Z_2$ time reversal symmetry breaking (QCP$_\text{Ising}$) and SU($2N$) flavor symmetry breaking (QCP$_\text{flavor}$) ground states, respectively.
On the kagome lattice, the former can be understood as chiral spin liquid, while the latter is expected to realize valence bond solid order.
The critical manifold with respect to time reversal symmetry breaking (flavor symmetry breaking) is indicated in blue (green).
For decreasing $N$, cQED$_3$ and QCP$_\text{flavor}$ approach each other and eventually collide at a critical flavor number $N_\text{c}$.
(b)~RG flow diagram for $N=7$ in the plane $e^2 = e^2_\star$ spanned by the four-fermion couplings $G_1$ and $G_2$, indicating the locations of the infrared stable cQED$_3$ fixed point and the quantum critical QCP$_\text{Ising}$ and QCP$_\text{flavor}$ fixed points.
(c)~Same as (b), but for the critical flavor number $N = N_\text{c} = 6.24$, indicating the fixed-point collision.
(d)~Same as (b), but for $N=2$, relevant for the extended Heisenberg spin-1/2 model on the kagome lattice. The absence of the cQED$_3$ fixed point implies that the U(1) Dirac spin liquid is unstable and the QCP$_\text{Ising}$ fixed point governs a direct continuous transition between valence bond solid order and a time-reversal-symmetry-breaking chiral spin liquid.
}
\label{fig:flow-diagram-QED3}
\end{figure*}

\subsection{Model}

Although the proposed mechanism for a continuous order-to-order transition does not depend on the formation of fractionalized quasiparticles, it is possible to emerge also in conjunction with fractionalization phenomena.
To illustrate this point, consider the extended Heisenberg model on the kagome lattice, defined by the Hamiltonian
\begin{align} \label{eq:kagome}
\mathcal H = J_1 \sum_{\langle ij \rangle} \vec S_i \cdot \vec S_j + J_2 \sum_{\llangle ij \rrangle} \vec S_i \cdot \vec S_j + J_3 \sum_{\lllangle ij \rrrangle} \vec S_i \cdot \vec S_j.
\end{align}
Here, $\vec S_i$ represents a spin-$1/2$ at site $i$, while ${\langle ij \rangle}$, ${\llangle ij \rrangle}$, and ${\lllangle ij \rrrangle}$ correspond to first-, \mbox{second-,} and third-neighbor bonds, respectively, with $J_1$, $J_2$, and $J_3$ denoting the associated exchange couplings, see Fig.~\ref{fig:kagome}(a).
The model describes magnetic Mott insulators crystallizing in a layered kagome structure, such as herbertsmithite ZnCu$_3$(OH)$_6$Cl$_2$~\cite{norman16}, kapellasite Cu$_3$Zn(OH)$_6$Cl$_2$~\cite{fak12}, or YCu$_3$(OH)$_{6.5}$Br$_{2.5}$~\cite{hong22, suetsugu24}.
For dominant antiferromagnetic $J_1 > 0$, numerical simulations at finite temperature or on finite lattices observe no evidence for any long-range-ordered state, suggesting the formation of a quantum spin liquid~\cite{ran07, yan11, depenbrock12, iqbal14, he17, liao17, chen18, schnack18, laeuchli19}.
The nature of this state has been a matter of significant debate, with recent simulations indicating a gapless U(1) Dirac spin liquid~\cite{he17, liao17, chen18}.
For increasing second- and third-neighbor interactions $J_2$ and $J_3$, a transition to a gapped chiral spin liquid, characterized by broken time reversal symmetry with finite scalar spin chirality $\langle \vec S_i \cdot (\vec S_j \times \vec S_k) \rangle$, has been found~\cite{he15a, he15b, hu15, gong15, wietek15}.
For instance, fixing the second- and third-nearest-neighbor couplings as $J_2/J_1 = J_3/J_1 = \alpha$ with antiferromagnetic next-neighbor coupling $J_1 > 0$, the transition occurs at $\alpha_\mathrm{c} \simeq 0.4$~\cite{he15a}.
The transition has been argued to be continuous~\cite{he15a} and described by the QED$_3$-Gross-Neveu model with $N=2$ four-component Dirac fields~\cite{he17, he15c, janssen17b, ihrig18, zerf18, boyack19, dupuis22}.
The QED$_3$-Gross-Neveu model describes fermionic excitations $\psi$ and $\psi^\dagger$ arising from fractionalization of the spin degrees of freedom, interacting via an emergent U(1) gauge field $a_\mu$ and additional local four-fermion interactions,
\begin{align}
    S_\text{QED$_3$-GN} & = \int \dd[3]{x} \Bigg[ \bar\psi_i \gamma_\mu \left(\partial_\mu - \frac{\rmi e}{\sqrt{N}} a_\mu \right)  \psi_i + \frac{1}{4} f_{\mu\nu}^2
     \nonumber\\&\quad    
    - \frac{G_1}{2N} \left( \bar\psi_i \gamma_{45} \psi_i \right)^2 - \frac{G_2}{2N} \left( \bar\psi_i \gamma_\mu \psi_i \right)^2 \Biggr]\,.
\end{align}
Here, $\psi_i$ and $\bar\psi_i \coloneqq \psi^\dagger \gamma_3$ denote $N=2$ flavors of four-component Dirac spinors in $(2+1)$-dimensional Euclidean space-time, $f_{\mu\nu} = \partial_\mu a_\nu - \partial_\nu a_\mu$ is the U(1) field-strength tensor associated with the U(1) gauge field $a_\mu$, the $4\times 4$ matrices $\gamma_\mu$ satisfy the Clifford algebra, and $\gamma_{45} \coloneqq \rmi \gamma_{4} \gamma_{5}$ is the Hermitian product of the two chiral gamma matrices in the reducible four-component spinor representation.
For notational simplicity, we have assumed the summation convention over repeated flavor indices $i = 1,\dots, N$ and space-time indices $\mu = 1,2,3$.
Among others, the model features $\mathbb Z_2$ time reversal and SU($2N$) flavor symmetry.

\subsection{RG analysis}

In 2+1 dimensions, the gauge coupling $e^2$ has positive mass dimension and flows to a finite value in the infrared. 
The four-fermion couplings $G_1$ and $G_2$ are power-counting irrelevant, but can be generated by the finite gauge coupling.
On the level of mean-field theory, the coupling $G_1$, if tuned beyond a certain finite threshold, induces a transition between the Dirac spin liquid for $G_1 < G_{1\text{c}}$ and the chiral spin liquid for $G_1 > G_{1\text{c}}$. The latter is characterized by a finite vacuum expectation value of the fermion bilinear $\langle \bar\psi_i \gamma_{45} \psi_i \rangle \neq 0$, which breaks time reversal symmetry~\cite{gies10}. The four-fermion coupling $G_1$ can thus be understood to parameterize the second- and third-nearest-neighbor interactions $J_2 = J_3$ in the microscopic model.
The coupling $G_2$ is generated by the RG flow and is therefore included from the outset.
Indeed, treating $G_1$ and $G_2$ on equal footing is essential for identifying the continuous order-to-order transition.

At one-loop order, the flow equations for the QED$_3$-Gross-Neveu model read~\cite{janssen16b,janssen17b}
\begin{align}
    \frac{\rmd e^2}{\rmd \ln b} & = \left(1 - \eta_a\right) e^2, 
    \allowdisplaybreaks[1] \\
    \frac{\rmd G_1}{\rmd \ln b} & = - G_1 + \frac{16}{3 N} e^2 G_1 + \frac{8}{N} e^2 G_2 - \frac{4}{N} e^4
    + \frac{2 (2N - 1)}{N} G_1^2 
    \nonumber\\&\quad
    - \frac{4}{N} G_2^2 - \frac{6}{N} G_1 G_2,
    \allowdisplaybreaks[1] \\
    \frac{\rmd G_2}{\rmd \ln b} & = - G_2 + \frac{8}{3 N} e^2 G_1 - \frac{2 (2N + 1)}{3 N} G_2^2
    - \frac{2}{N} G_1 G_2,
\end{align}
with gauge field anomalous dimension $\eta_a = \frac{4}{3} e^2$.
Above, we employed the rescaling $e^2 \mapsto 2 \pi^2 \Lambda e^2$, $G_1 \mapsto 2 \pi^2 \Lambda^{-1} G_1$, and $G_2 \mapsto 2 \pi^2 \Lambda^{-1} G_2$, where $\Lambda$ denotes the ultraviolet cutoff.
Figure~\ref{fig:flow-diagram-QED3}(a) illustrates the fixed-point structure of the model.
For sufficiently large $N$, the infrared attractive plane $e^2 = e^2_\star = 3/4$ hosts a fully infrared stable fixed point (cQED$_3$), associated with the conformal phase of QED$_3$~\cite{braun14, dipietro16, janssen16b, herbut16a}.
In the context of the extended Heisenberg spin-1/2 model on the kagome lattice, the latter can be understood as a U(1) Dirac spin liquid~\cite{he17, liao17, chen18}.
In addition, the plane features two quantum critical fixed points, associated with instabilities towards Ising $\mathbb Z_2$ time reversal symmetry breaking (QCP$_\text{Ising}$) and SU($2N$) flavor symmetry breaking (QCP$_\text{flavor}$) ground states, respectively~\cite{braun14,herbut16a}.
On the kagome lattice, the former can be understood as chiral spin liquid~\cite{he17, he15c}, while the latter is expected to realize valence bond solid order~\cite{hermele05}.

\begin{figure}[tb!]
\includegraphics[width=\linewidth]{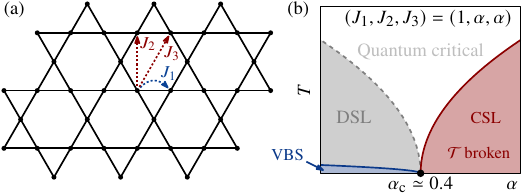}
\caption{%
(a)~Kagome lattice, indicating first ($J_1$), second ($J_2$), and third ($J_3$) neighbors.
(b)~Schematic finite-temperature phase diagram of kagome Heisenberg model as function of $\alpha = J_2/J_1 = J_3/J_1$, from RG analysis of effective QED$_3$-Gross-Neveu model.
The gapped chiral spin liquid (CSL) breaks time reversal. 
Solid (dashed) lines denote phase transitions (crossovers).
The disordered gapless Dirac spin liquid (DSL) is unstable at low temperatures towards a valence bond solid (VBS), leaving behind a continuous zero-temperature transition between distinct orders.}
\label{fig:kagome}
\end{figure}

For decreasing $N$, the infrared stable cQED$_3$ fixed point and the quantum critical QCP$_\text{flavor}$ fixed point approach each other and eventually collide at a critical flavor number $N_\text{c}$.
This is illustrated in Figs.~\ref{fig:flow-diagram-QED3}(b)--(d), which shows the RG flow in the $G_1$-$G_2$ plane for fixed $e^2 = e^2_\star$ and different values of $N$ above, at, and below the critical flavor number $N_\text{c}$.
For $N < N_\text{c}$, the cQED$_3$ and QCP$_\text{flavor}$ fixed points are located in the complex plane, leaving behind a runaway flow in the space of real couplings.
Importantly, the fixed point QCP$_\text{Ising}$, associated with the time-reversal-symmetry-breaking instability, remains unaffected by the collision and annihilation process. For $N < N_\text{c}$, it distinguishes between two different regimes characterized by runaway flows: Consider the line of starting values given by $e^2 = e^2_\star$ and $G_2 = 0$ at the ultraviolet scale. For $G_1 > G_{1\star}$, the flow diverges towards $(G_1,G_2) \to (\infty,0)$ at a finite RG scale, indicating spontaneous time reversal symmetry breaking. For $G_1 < G_{1\star}$, on the other hand, the flow diverges towards $(G_1, G_2) \to (-\infty, -\infty)$, which has previously been interpreted as an instability towards flavor symmetry breaking~\cite{gies10,janssen17b,herbut16a}.
This suggests that QCP$_\text{Ising}$ governs for $N < N_\text{c}$ a continuous quantum phase transition on the  between two long-range ordered states that break different symmetries. 
At the one-loop order, we find the critical flavor number as $N_\text{c} = 6.24$. Higher-loop contributions may reduce the number, but most analytical estimates~\cite{braun14, dipietro16, herbut16a, gusynin16, kotikov16} appear to suggest that the fixed-point collision and annihilation occurs \emph{above} the physical number $N=2$ relevant for the kagome quantum magnets. (We emphasize, however, that we are not aware of clear numerical evidence for the scenario in an unbiased simulation of the QED$_3$-Gross-Neveu model, such as Monte Carlo~\cite{karthik16a, karthik16b}.)
If indeed $N_\text{c} > 2$ in the QED$_3$-Gross-Neveu model, it implies for the extended Heisenberg spin-1/2 model on the kagome lattice that 
(1)~the U(1) Dirac spin liquid becomes unstable at the lowest temperatures towards a valence bond solid ordered ground state and
(2)~the quantum phase transition between the valence bond solid, stabilized for $\alpha < 0.4$, and the chiral spin liquid, stabilized for $\alpha > 0.4$, is continuous and governed by the QCP$_\text{Ising}$ fixed point.
The resulting schematic phase diagram is depicted in Fig.~\ref{fig:kagome}(b).
An explicit computation of the order parameters across the order-to-order transition would require introducing the corresponding Hubbard-Stratonovich fields within the dynamical bosonization framework of Sec.~\ref{sec:pyrochlore-iridates}, which we leave for future work.

\section{Conclusions}
\label{sec:conclusions}

We have proposed a mechanism for generic continuous order-to-order quantum phase transitions that operates independently of fractionalization.
This mechanism is based on the collision and annihilation of two RG fixed points, and may be realized in pyrochlore iridates and kagome quantum magnets.
Further possible realizations in quantum impurity models and quantum chromodynamics with additional four-fermion interactions are discussed in \ref{app:further-examples}.
A distinctive experimental or numerical signature of the mechanism is a pronounced asymmetry in the effective energy scales between the two sides of the transition, characterized by a significantly lower critical temperature and a much smaller order parameter on the side where the order emerges from fixed-point annihilation.

The fixed-point annihilation mechanism responsible for the continuous order-to-order transition, discussed in this work, should be contrasted with previously discussed scenarios, where fixed-point annihilation gives rise to weak first-order transitions~\cite{halperin74,gukov17,gorbenko18a,hawashin25}. Such a scenario has recently been discussed in the context of SU($N$) quantum spin models on two-dimensional lattices, which are believed to be described by the $N$-component Abelian Higgs model~\cite{block13,nahum15b,ihrig19,senthil23,song25}.
Similar scenarios have also been discussed in $Q$-state Potts models~\cite{gorbenko18b, ma19, jacobsen24, tang24} and tensor O$(N)$ models~\cite{fei14,herbut16b,gracey18c,janssen22}.
In all these examples, a critical fixed point collides with a bicritical fixed point, and both disappear into the complex plane as a function of some external parameter.
In these systems, after the fixed-point collision, a continuous order-to-order transition is therefore not possible without additional fine tuning.
In the examples discussed in the present work, by contrast, a critical fixed point collides with an infrared stable fixed point, destabilizing the symmetric phase and leaving behind a continuous transition between two different orders.

\section*{Acknowledgements}
%
We thank Aiman Al-Eryani, Igor Boettcher, Rufus Boyack, Shankar Ganesh, Bilal Hawashin, Igor F.\ Herbut, Joseph Maciejko, Zi~Yang Meng, Max A.\ Metlitski, Adam Nahum, Shouryya Ray, Michael M.\ Scherer, Carsten Timm, Mireia Tolosa-Simeón, Matthias Vojta, and Manuel Weber for illuminating discussions.
This work has been supported by the Deutsche Forschungsgemeinschaft (DFG) through 
Project No.~247310070 (SFB 1143, Project A07), 
Project No.~390858490 (W\"{u}rzburg-Dresden Cluster of Excellence {\it ct.qmat}, EXC 2147), and 
Project No.~411750675 (Emmy Noether program, JA2306/4-1).

\section*{Data availability}
%
The data that support the findings of this article are openly available~\cite{data-availability}.

\appendix

\section{Solid-angle integrals}
\label{app:solid-angle-integrals}

\begin{table*}[bt!]
    \caption{Values of solid-angle integrals $f_i(\delta)$ for limiting cases $\delta = \pm 1$ and isotropic case $\delta = 0$ from, whenever possible, analytical integration, otherwise determined numerically.}
    \begin{subtable}[c]{0.27\textwidth}
    \begin{tabular*}{\linewidth}{@{\extracolsep{\fill} } c c c c }
    \hline\hline
        $f_i$ & $f_i (-1)$ & $f_i(0)$ & $f_i (+1)$ \\ \hline
        $f_{1}$ & 1.0942 & 1 & 0.8130 \\
        $f_{1 \text{t}}$ & 1.3294 & 1 & 0.6225 \\
        $f_{1 \text{e}}$ & 0.7415 & 1 & 1.0987 \\
        $f_{2}$ & 1 & 1 & 1/2 \\
        $f_{2 \text{t}}$ & 5/6 & 1 & 0.6775 \\
        $f_{2 \text{e}}$ & 1.3678 & 1 & 5/8 \\
        $f_{3}$ & 4/3 & 1 & 2/3 \\
    \hline\hline
    \end{tabular*}
    \end{subtable}
    \hfill
    \begin{subtable}[c]{0.27\textwidth}
    \centering
    \begin{tabular*}{\linewidth}{@{\extracolsep{\fill} } c c c c }
    \hline\hline
        $f_i$ & $f_i (-1)$ & $f_i(0)$ & $f_i (+1)$ \\ \hline
        $f_{3 \text{t}}$ & 5/9 & 1 & 5/9 \\
        $f_{3 \text{e}}$ & 5/3 & 1 & 5/12 \\
        $f_{3 \overline{\text{t}}}$ & 35/54 & 1 & 0.7262 \\
        $f_{4}$ & 16/5 & 1 & 8/5 \\
        $f_{4 \text{t}}$ & 4/3 & 1 & 4/9 \\
        $f_{4 \text{e}}$ & 4/3 & 1 & 1 \\
        $f_{4 \text{t} \text{t}}$ & 7/27 & 1 & 7/9 \\
    \hline\hline
    \end{tabular*}
    \end{subtable}
    \hfill
    \begin{subtable}[c]{0.27\textwidth}
    \begin{tabular*}{\linewidth}{@{\extracolsep{\fill} } c c c c }
    \hline\hline
        $f_i$ & $f_i (-1)$ & $f_i(0)$ & $f_i (+1)$ \\ \hline
        $f_{4 \text{t} \text{t}'}$ & 7/9 & 1 & 7/18 \\
        $f_{4 \text{e} \text{e}}$ & 7/3 & 1 & 7/16 \\
        $f_{4 \text{e} \text{t}}$ & 7/9 & 1 & 7/24 \\
        $f_{4 \text{t} \overline{\text{t}}}$ & 77/162 & 1 & 0.8876 \\
        $f_{4 \text{e} \overline{\text{t}}}$ & 77/108 & 1 & 77/144 \\
        $f_{e^2}$ & 16/9 & 1 & 4/3 \\
        $f_{g^2}$ & 16/9 & 0 & $-2/3$ \\
    \hline\hline
    \end{tabular*}
    \end{subtable}
    \label{TablefiLimits}
\end{table*}

In this appendix, we provide definitions and numerical values of the dimensionless functions $f_i(\delta)$, with $i \in \{1$, $1\text{t}$, $1\text{e}$, $2$, $2\text{t}$, $2\text{e}$, $3$, $3\text{t}$, $3\text{e}$, $3\overline{\text{t}}$, $4$, $4\text{t}$, $4\text{e}$, $4\text{tt}$, $4\text{tt}'$, $4\text{ee}$, $4\text{et}$, $4\text{t}\overline{\text{t}}$, $4\text{e}\overline{\text{t}}, e^2\}$, and $f_{g^2}(\delta)$ as function of the anisotropy parameter $\delta$, occurring in the mean-field calculation in Sec.~\ref{subsec:mft} and in the RG analysis in Sec.~\ref{subsec:rg}.
These are bounded and continuous functions of order unity with $f_i >0$ ($f_{g^2} \geq -2/3$) for $\delta \in [-1,1]$ and $f_i = 1$ ($f_{g^2} = 0$) for $\delta =0$.
Some of these have already been defined in Ref.~\cite{boettcher17}, reading
\begin{align}
    f_{1} (\delta) &\coloneqq \frac{1}{4 \pi} \int \dd{\Omega} \frac{1}{\tilde{X}^{1/2}}, \allowdisplaybreaks[1] \\
    f_{1 \text{t}} (\delta) &\coloneqq \frac{5}{4 \pi} \int \dd{\Omega} \frac{\tilde{d}_1^2}{\tilde{X}^{1/2}}, \allowdisplaybreaks[1] \\
    f_{1 \text{e}} (\delta) &\coloneqq \frac{5}{4 \pi} \int \dd{\Omega} \frac{\tilde{d}_4^2}{\tilde{X}^{1/2}}, \allowdisplaybreaks[1] \\
    f_{2} (\delta) &\coloneqq \frac{1}{4 \pi} (1-\delta) (1+\delta) \int \dd{\Omega} \frac{1}{\tilde{X}^{3/2}}, \allowdisplaybreaks[1] \\
    f_{2 \text{t}} (\delta) &\coloneqq \frac{5}{4 \pi} (1+\delta) \int \dd{\Omega} \frac{\tilde{d}_1^2}{\tilde{X}^{3/2}}, \allowdisplaybreaks[1] \\
    f_{2 \text{e}} (\delta) &\coloneqq \frac{5}{4 \pi} (1-\delta) \int \dd{\Omega} \frac{\tilde{d}_4^2}{\tilde{X}^{3/2}}, \allowdisplaybreaks[1] \\
    f_{3} (\delta) &\coloneqq \frac{1}{4 \pi} (1-\delta)^3 (1+\delta)^3 \int \dd{\Omega} \frac{1}{\tilde{X}^{5/2}}, \allowdisplaybreaks[1] \\
    f_{3 \text{t}} (\delta) &\coloneqq \frac{5}{4 \pi} (1-\delta) (1+\delta)^3 \int \dd{\Omega} \frac{\tilde{d}_1^2}{\tilde{X}^{5/2}}, \allowdisplaybreaks[1] \\
    f_{3 \text{e}} (\delta) &\coloneqq \frac{5}{4 \pi} (1-\delta)^3 (1+\delta) \int \dd{\Omega} \frac{\tilde{d}_4^2}{\tilde{X}^{5/2}}, \allowdisplaybreaks[1] \\
    f_{3 \overline{\text{t}}} (\delta) &\coloneqq \frac{35}{\sqrt{3} 4 \pi} (1+\delta)^3 \int \dd{\Omega} \frac{\tilde{d}_1 \tilde{d}_2 \tilde{d}_3}{\tilde{X}^{5/2}},
\end{align}
where $\int \dd{\Omega} = \int_0^\pi \dd{\theta} \sin \theta \int_0^{2 \pi} \dd{\phi}$ denotes the integration over the solid angle, $\tilde{X} (\theta, \phi) = (1-\delta)^2 + 12 \delta \sum_{i<j} \frac{q_i^2}{q^2} \frac{q_j^2}{q^2}$, and $\tilde{d}_a (\theta, \phi) \coloneqq d_a (\vec{q}) / q^2$ are the $\ell = 2$ real spherical harmonics.
Upon spatial rotations, the latter transform under the irreducible representation $\text{T}_{2 \text{g}}$ ($\text{E}_\text{g}$) of the octahedral point group $\text{O}_\text{h}$ for $a=1,2,3$ ($a=4,5$), and the indices t and e of the functions $f_i$ indicate the type of spherical harmonics involved in the integral.
In addition to the above-defined functions, the loop expansion of the AIAO order-parameter field theory~\cite{moser24} gave rise to the following new solid-angle integrals,
\begin{align}
    f_{4} (\delta) &\coloneqq \frac{1}{4 \pi} (1-\delta)^5 (1+\delta)^5 \int \dd{\Omega} \frac{1}{\tilde{X}^{7/2}},\\
    f_{4 \text{t}} (\delta) &\coloneqq \frac{5}{4 \pi} (1-\delta)^3 (1+\delta)^5 \int \dd{\Omega} \frac{\tilde{d}_1^2}{\tilde{X}^{7/2}}, \allowdisplaybreaks[1] \\
    f_{4 \text{e}} (\delta) &\coloneqq \frac{5}{4 \pi} (1-\delta)^5 (1+\delta)^3 \int \dd{\Omega} \frac{\tilde{d}_4^2}{\tilde{X}^{7/2}}, \allowdisplaybreaks[1] \\
    f_{4 \text{t} \text{t}} (\delta) &\coloneqq \frac{35}{12 \pi} (1-\delta) (1+\delta)^5 \int \dd{\Omega} \frac{\tilde{d}_1^2 \cdot \tilde{d}_1^2}{\tilde{X}^{7/2}}, \allowdisplaybreaks[1] \\
    f_{4 \text{t} \text{t}'} (\delta) &\coloneqq \frac{35}{4 \pi} (1-\delta) (1+\delta)^5 \int \dd{\Omega} \frac{\tilde{d}_1^2 \cdot \tilde{d}_2^2}{\tilde{X}^{7/2}}, \allowdisplaybreaks[1] \\
    f_{4 \text{e} \text{e}} (\delta) &\coloneqq \frac{35}{12 \pi} (1-\delta)^5 (1+\delta) \int \dd{\Omega} \frac{\tilde{d}_4^2 \cdot \tilde{d}_4^2}{\tilde{X}^{7/2}}, \allowdisplaybreaks[1] \\
    f_{4 \text{e} \text{t}} (\delta) &\coloneqq \frac{35}{4 \pi} (1-\delta)^3 (1+\delta)^3 \int \dd{\Omega} \frac{\tilde{d}_1^2 \cdot \tilde{d}_4^2}{\tilde{X}^{7/2}}, \allowdisplaybreaks[1] \\
    f_{4 \text{t} \overline{\text{t}}} (\delta) &\coloneqq \frac{385}{\sqrt{3} 12 \pi} (1+\delta)^5 \int \dd{\Omega} \frac{\tilde{d}_1^2 \cdot \tilde{d}_1 \tilde{d}_2 \tilde{d}_3}{\tilde{X}^{7/2}}, \allowdisplaybreaks[1] \\
    f_{4 \text{e} \overline{\text{t}}} (\delta) &\coloneqq \frac{385}{\sqrt{3} 4 \pi} (1-\delta) (1+\delta)^3 \int \dd{\Omega} \frac{\tilde{d}_4^2 \cdot \tilde{d}_1 \tilde{d}_2 \tilde{d}_3}{\tilde{X}^{7/2}}.
\end{align}
For the Coulomb and AIAO-order-parameter anomalous dimensions, it is convenient to define two further functions as combinations of the above-defined ones,
\begin{align}
    f_{e^2} (\delta) &\coloneqq \frac{1}{3} \left[ 2 (1 - \delta)^2 + 3 (1 + \delta)^2 \right] f_2 (\delta)
    \nonumber\\&\quad
    - \frac{2}{3} \bigg[ \frac{2}{5} (1 - \delta)^2 f_{3 \text{e}} (\delta) + \frac{3}{5} (1 + \delta)^2 f_{3 \text{t}} (\delta)
    \nonumber\\&\quad
    - \frac{12}{5} \frac{\delta^2}{(1 + \delta)^2} f_{3 \text{t}} (\delta) + \frac{36}{35} \frac{\delta^2 (1 - \delta)}{(1 + \delta)^2} f_{3 \overline{\text{t}}} (\delta) \bigg],
    \allowdisplaybreaks[1] \\
    f_{g^2} (\delta) &\coloneqq \frac{2 (1 - \delta)^2 - (1 + \delta)^2}{4} f_2(\delta)
    \nonumber\\&\quad
    + \frac{8 \delta^2 - 6 (1 - \delta)^2 (1 + \delta)^2 - 11 (1 + \delta)^4}{20 (1 + \delta)^2} f_{3 \text{t}}(\delta)\nonumber\\&\quad
    + \frac{4 (1 - \delta)^2 + 9 (1 + \delta)^2}{30} f_{3 \text{e}}(\delta) - \frac{6 \delta^2 (1 - \delta)}{35 (1 + \delta)^2} f_{3 \overline{\text{t}}}(\delta)
    \allowdisplaybreaks[1] \nonumber\\&\quad
    + \Bigg\{ \frac{(1 - \delta)^2 \left[-4 (1 - \delta)^2 + 7 (1 + \delta)^2\right]}{14 (1 + \delta)^2}
    \nonumber\\&\quad\qquad
    + \frac{2 \delta^2 \left[-(1 - \delta)^2 + 9 (1 + \delta)^2\right]}{7 (1 + \delta)^4} \Bigg\} f_{4 \text{t} \text{t}}(\delta)
    \allowdisplaybreaks[1] \nonumber\\&\quad
    + \Bigg\{ \frac{(1 - \delta)^2 \left[-5 (1 - \delta)^2 + 
   11 (1 + \delta)^2\right]}{42 (1 + \delta)^2}
   \nonumber\\&\quad\qquad
    + \frac{2 \delta^2 \left[(1 - \delta)^2 + 18 (1 + \delta)^2\right]}{21 (1 + \delta)^4} \Bigg\} f_{4 \text{t} \text{t}'}(\delta)
    \allowdisplaybreaks[1] \nonumber\\&\quad
    - \frac{4 (1 - \delta)^2}{21}  f_{4 \text{e} \text{e}}(\delta) + \frac{54 \delta^2 (1 - \delta)}{77 (1 + \delta)^2} f_{4 \text{t} \overline{\text{t}}}(\delta)
    \allowdisplaybreaks[1] \nonumber\\&\quad
    - \frac{4 \delta \left[(1 - \delta)^2 + 3 (1 + \delta)^2\right]}{21 (1 + \delta)^2} f_{4 \text{e} \text{t}}(\delta)
    \allowdisplaybreaks[1] \nonumber\\&\quad
    - \frac{12 \delta^2 (1 - \delta)^2}{77 (1 + \delta)^2} f_{4 \text{e} \overline{\text{t}}}(\delta).
\end{align}
%
%
\begin{figure}[tb!]
\centering
\includegraphics[width=\linewidth]{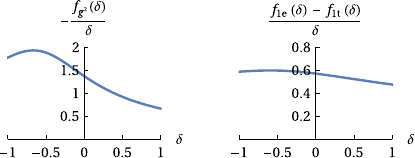}
\caption{%
Graphs of two useful expression, $-\frac{f_{g^2}}{\delta}$ (left) and $\frac{f_{1 \text{e}} - f_{1 \text{t}}}{\delta}$ (right), arising frequently in the RG analysis, as function of anisotropy parameter $\delta$. Most importantly, both functions only possess a removable singularity at $\delta = 0$.}
\label{fig:fi-quotient}
\end{figure}
%
We emphasize that all $f_i$ are bounded from above and below for all $\delta \in [-1,1]$; in particular, they remain finite in the limiting cases $\delta = \pm 1$, see Table~\ref{TablefiLimits}.
For graphs of the above defined functions $f_i$, we refer the reader to Refs.~\cite{boettcher17,moser24}.
Further, the two ratios, $-\frac{f_{g^2} (\delta)}{\delta}$ and $\frac{f_{1 \text{e}} (\delta) - f_{1 \text{t}} (\delta)}{\delta}$, arise frequently in the RG analysis in Sec.~\ref{subsec:rg}.
Their graphs are displayed in Fig.~\ref{fig:fi-quotient}.
Most importantly, both ratios are well defined in the isotropic limit $\delta \rightarrow 0$, namely $\lim_{\delta \rightarrow 0} \frac{f_{g^2} (\delta)}{-\delta} = \frac{48}{35}$ and $\lim_{\delta \rightarrow 0} \frac{f_{1 \text{e}} (\delta) - f_{1 \text{t}} (\delta)}{\delta} = \frac{4}{7}$.

\section{Fermionic RG flow}
\label{app:fermionic-flow}

\begin{figure*}[t!]
\includegraphics[width=\linewidth]{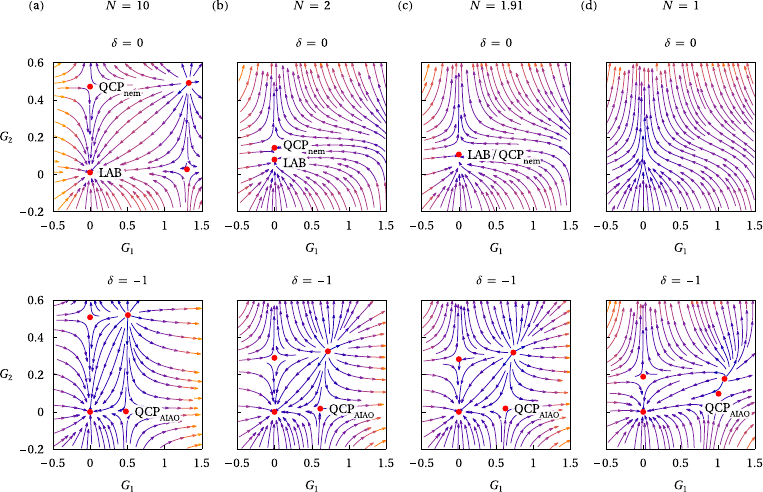}
\caption{%
(a)~RG flow diagram in $d=3$ spatial dimensions and for $N=10$ in the point-like limit $\alpha_1, \alpha_2 \to 1$ [Eqs.~\eqref{eq:delta-point-like}--\eqref{eq:eta-a-point-like}] in the plane spanned by $G_1$ and $G_2$ for fixed $(\delta, e^2) = (0, 15N/(4+15N))$ (top panel) and $(\delta, e^2) = (-1, 9/16)$ (bottom panel), respectively.
Arrows indicate flow towards infrared.
Besides the infrared stable fixed point associated with the disordered LAB phase (LAB), there are two quantum critical points associated with instabilities towards nematic (QCP$_\text{nem}$) and AIAO (QCP$_\text{AIAO}$) orders, respectively.
(b)~Same as (a), but for $N=2$, illustrating that the LAB and QCP$_\text{nem}$ fixed point approach each other for decreasing $N$.
(c)~Same as (a), but for $N = 1.91$, illustrating the collision of the LAB and QCP$_\text{nem}$ fixed points at a critical $N>1$.
(d)~Same as (a), but for $N = 1$, illustrating the runaway flow in the isotropic plane $\delta = 0$ (top panel) and the persistence of QCP$_\text{AIAO}$ in the plane $\delta = -1$ (bottom panel).
}
\label{fig:fermionic-flow-diagram}
\end{figure*}

In this appendix, we show that the dynamically bosonized flow equations of our model using Hubbard-Stratonovich fields $\phi$ and $\varphi$, as obtained in Sec.~\ref{subsec:rg}, reduce, in the limit of infinite order-parameter masses, $\alpha_1, \alpha_2 \to 1$, to those of a purely fermionic model, obtained without introducing order-parameter fields explicitly. The latter is defined by the action
\begin{align}
    S_\text{F} & = \int \dd{\tau} \dd[3]{\vec x} \Biggl\{
    \psi^{\dagger} \left( \partial_\tau + \sum_{a = 1}^5 (1 + s_a \delta) d_a \gamma_a + \frac{\rmi e}{\sqrt{N}} a \right) \psi	
    \nonumber \\ & \quad
     + \frac{1}{2} (\nabla a)^2
    - \frac{G_1}{2N} (\psi^\dagger \gamma_{45} \psi )^2
    - \frac{G_2}{2N} \sum_{a=1}^5 (\psi^\dagger \gamma_{a} \psi)^2
    \Biggr\}\,,
    \label{eq:Luttinger-four-fermion-model}
\end{align}
and has been studied previously in Refs.~\cite{herbut14,janssen17a,boettcher17}.
The goal of the comparison is twofold.
First, we want to demonstrate that our calculations are consistent with the fermionic ones presented in Refs.~\cite{herbut14,janssen17a,boettcher17}.
Second, we want build some intuition on fermionic flows with respect to the proposed fixed-point annihilation mechanism, which prepares us for the examples discussed below.
We note, however, that the limit of infinite order-parameter masses is in principle insufficient to compute order parameters in symmetry-broken phases, as required for explicitly demonstrating the proposed order-to-order transition. This is because fermionic RG flows diverge in the case of spontaneous symmetry breaking.

We start from the flow equations in the dynamical bosonization scheme in terms of the couplings $\delta$, $e^2$, $G_1$, $\alpha_1$, $G_2$, $\lambda$, $c$, and $\alpha_2$.
In the limit of infinite order-parameter masses, $\alpha_1, \alpha_2 \to 1$, the flow equations for $\delta$, $e^2$, $G_1$, and $G_2$ become independent of $\lambda$ and $c$. In this limit, we find
\begin{align} \label{eq:delta-point-like}
    \frac{\rmd\delta}{\rmd \ln b} & = - \eta_\psi \delta + \frac{2}{15 N} (1-\delta^2) [ (1 + \delta) f_{1 \text{t}} - (1 - \delta) f_{1 \text{e}} ] e^2,
\allowdisplaybreaks[1] \\
    \frac{\rmd e^2}{\rmd \ln b} & = \left(z+2-d- \eta_a + \frac{2\delta}{1-\delta^2}\frac{\rmd \delta}{\rmd \ln b}\right) e^2, 
\allowdisplaybreaks[1] \\
    \frac{\rmd G_1}{\rmd \ln b} & = (z - d) G_1 + \frac{2}{5 N} (1-\delta) (1-\delta^2) f_{2 \text{e}} e^2 G_1
    \nonumber\\&\quad
    - \frac{2 ( 1 - 2 N )}{5 N} (1-\delta) f_{2 \text{e}} G_1^2
    \nonumber\\&\quad
    - \frac{2}{5 N} (1-\delta) f_{2 \text{e}} G_1 G_2,
\allowdisplaybreaks[1] \\
    \frac{\rmd G_2}{\rmd \ln b} & = (z^{(0)} - d) G_2 + \frac{37 + 16 N}{10 N} G_2^2 + \frac{4}{5 N} (1-\delta^2) e^2 G_2
    \nonumber\\&\quad
    + \frac{1}{10 N} (1-\delta^2)^2 e^4 - \frac{4 c_1 + 5 c_2}{5 N} G_1 G_2
    \nonumber\\&\quad
    + \frac{1}{10 N} G_1^2,
\end{align}
with the anomalous dimensions
\begin{align}
    \eta_\psi & = \frac{2}{15 N} (1-\delta^2) [(1+\delta) f_{1 \text{t}} + (1-\delta) f_{1 \text{e}}] e^2, \allowdisplaybreaks[1] \\
    \eta_a & = e^2 f_{e^2}, \label{eq:eta-a-point-like}
\end{align}
the full anisotropy-dependent dynamical exponent $z = 2 - \eta_\psi$, and the corresponding form $z^{(0)} = 2- \frac{4}{15 N} (1-\delta^2) e^2$ in the isotropic limit $\delta \to 0$, relevant for the flow in the nematic channel $G_2$.
To arrive at the above flow equations, we have rescaled the AIAO order-parameter mass in the dynamically bosonized flow equations as $r_1 \mapsto (1-\delta^2) r_1$, such that $G_1 = g_1^2/r_1$ remains finite at the AIAO quantum critical point.
The dimensionless couplings occurring in Eqs.~\eqref{eq:delta-point-like}--\eqref{eq:eta-a-point-like} are then related to the dimensionful couplings of the microscopic action, Eq.~\eqref{eq:Luttinger-four-fermion-model}, via the rescaling $e^2 \mapsto 2\pi^2 \Lambda^{4-d} (1-\delta^2)e^2$, $G_1 \mapsto 2 \pi^2 \Lambda^{2-d} G_1$, and $G_2 \mapsto 2 \pi^2 \Lambda^{2-d} G_2$.
For $N=1$, the flow equations for the anisotropy $\delta$, the charge $e^2$, and the anomalous dimensions $\eta_\psi$ and $\eta_a$ agree with the previous calculation~\cite{boettcher17}.
For all $N \in \mathbb N$, but $\delta = 0$, the flow equations for the charge $e^2$ and the anomalous dimensions $\eta_\psi$ and $\eta_a$ agree with those of Refs.~\cite{herbut14,janssen17a}. The flow of $G_2$ also agrees, with the exception of slight differences in the prefactors of the terms $\propto G_1 G_2$ and $\propto G_1^2$. These can be attributed to the fact that we have performed the dynamical bosonization scheme only in the nematic channel, leading to a Fierz-incomplete calculation.
This also explains the (minor) differences in the flow of $G_1$ in comparison with the corresponding calculation in Refs.~\cite{herbut14, janssen17a}.

For $N > 1.91$, the fermionic flow equations feature a fully infrared stable fixed point, associated with the LAB phase, as well as two quantum critical fixed points, associated with the instabilities towards nematic and AIAO orders, respectively.
We note that the AIAO fixed point in the point-like limit, in contrast to our results in the dynamical bosonization scheme (Sec.~\ref{subsubsec:fixed-point-structure}), exhibits an additional marginally relevant direction, associated with the flow of the anisotropy parameter $\delta$. This is a known artifact of the fermionic RG flow that can be attributed to the negligence of AIAO order-parameter fluctuations in the point-like limit~\cite{moser24}.
For $N = 1.91$, the infrared stable LAB fixed point and the nematic quantum critical point collide and subsequently disappear into the complex plane for $N < 1.91$, in qualitative agreement with the fixed-point structure obtained from the dynamically bosonized RG flow, discussed in Sec.~\ref{subsec:rg}, as well as the earlier works~\cite{herbut14,janssen17a,boettcher17}.
In agreement with the dynamically bosonized RG flow, the fixed-point collision takes place in the isotropic $G_1$-$G_2$ plane defined by $\delta = 0$ and finite effective charge $e^2 = \frac{15 N}{4 + 15 N}$. The flow in this plane is depicted for different values of $N$ in Fig.~\ref{fig:fermionic-flow-diagram} (top row), illustrating the fixed-point collision.
Although all fixed points in this section are analytically accessible, we refrain from stating their expressions here, as they are rather lengthy.

Importantly, the AIAO fixed point does not take part in the collision and annihilation of the LAB fixed point with the nematic fixed point, but remains in the physical space of real couplings for all $N \in \mathbb N$, in agreement with our results in the dynamical bosonization scheme. 
The AIAO fixed point is located in the plane defined by anisotropy $\delta = -1$ and effective charge $e^2 = \frac{9}{16}$. The flow in this plane is depicted for different values of $N$ in Fig.~\ref{fig:fermionic-flow-diagram} (bottom row), illustrating the fact that the AIAO fixed point persists for all $N \in \mathbb N$.
We conclude, that, at least in the case of Luttinger semimetals, the presented fixed-point annihilation mechanism is also available in the purely fermionic formulation.

\section{Higher-loop corrections}
\label{app:higher-loop-corrections}

In this appendix, we demonstrate that the critical exponents at the AIAO fixed point we have found at the one-loop order do not receive higher-loop corrections, and are thus expected to be exact.
This is ultimately another consequence of the maximal anisotropy $\delta_\star = -1$ at the AIAO fixed point.
Our argument is independent of the number of quadratic band touching points $N$ and therefore holds for all $N \in \mathbb N$.

As argued in Ref.~\cite{moser24} for $G_2 =0$, higher-loop corrections in the AIAO and charge sector are fully suppressed at the AIAO fixed point. This can be understood as a consequence of the fact that the AIAO fixed point is located at maximal anisotropy $\delta_\star = -1$.
Here, it can be seen from the fact that the fixed point values for $e^2$ and $g_1^2$ are finite \emph{after} performing the anisotropy-dependent reparametrization $(e^2,g_1^2) \mapsto (1 - \delta^2) (e^2,g_1^2)$, cf.~Sec.~\ref{subsec:rg}.
After this rescaling, every higher-loop diagram now comes with a strictly positive power of $(1 - \delta^2)$.
As a result, all higher-loop corrections are strongly suppressed in the vicinity of the AIAO fixed point at anisotropy $\delta_\star = -1$, rendering the one-loop flow equations asymptotically exact, see Ref.~\cite{moser24} for details and Refs.~\cite{huh08, schwab22} for a different model with analogous behavior.

\begin{figure}[tb]
\centering
\includegraphics[width=0.72\linewidth]{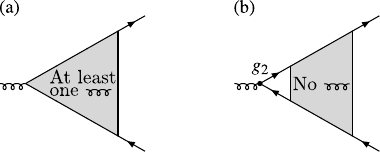}
\caption{%
Every loop contribution to the nematic Yukawa vertex falls into one of two distinct classes, (a) the class of diagrams which involve at least one internal nematic boson and (b) the class of diagrams which do not involve any internal nematic boson.
While class (a) does not contribute at the AIAO fixed point due to infinite nematic boson mass $r_{2\star} = \infty$, class (b) vanishes at the AIAO fixed point due to maximal spatial anisotropy $\delta_\star = -1$.}
\label{fig:diagram-nematic-yukawa-vertex-higher-loop}
\end{figure}

In our extended model for finite $G_1$ and $G_2$, the AIAO fixed point remains to be located at maximal anisotropy $\delta_\star = -1$ and additionally lies now at infinite nematic boson mass $r_{2\star} \to \infty$, cf.\ Table~\ref{TableAIAOQCP}.
These two fixed-point values suppress all higher-loop diagrams contributing to the flow equations relevant for the critical exponents.
For instance, all higher-loop corrections to the flow of the anisotropy $\delta$, to the effective charge $e^2$, and to the AIAO Yukawa coupling $g_1^2$ are either strongly suppressed by strictly positive powers of $(1-\delta^2)$ (these are the already known contributions arising in the theory for $G_1 = 0$) and/or by strictly positive powers of $1/(1 + r_2)$ (these are new contributions with at least one internal nematic boson) in the vicinity of the AIAO fixed point.
Consequently, the anomalous dimension of the fermion $\eta_\psi = 0$, the dynamical exponent $z = 2 - \eta_\psi = 2$, the anomalous dimension of the photon $\eta_a = 4 - d$, and the anomalous dimension the AIAO boson $\eta_\phi = 4 - d$ are exact critical exponents of the AIAO fixed point.
A similar argument can be made for the new flow equation of the nematic Yukawa coupling $g_2^2$ here, since all loop-correction fall again into two distinct classes, as depicted in Fig.~\ref{fig:diagram-nematic-yukawa-vertex-higher-loop}, which both vanish at the AIAO fixed point.
The first class of diagrams consists of all loop-corrections involving at least one internal nematic boson.
Since internal nematic bosons give rise to the appearance of strictly positive powers of $1/(1 + r_2)$, the whole first class of diagrams is suppressed in the vicinity of the AIAO fixed point located at infinite nematic mass $r_{2\star} = \infty$.
Any higher-loop order correction that is not of the above form involves only internal fermion, photon, and/or AIAO boson propagators and is, following the argument presented in Ref.~\cite{moser24}, expected to be suppressed by a strictly positive power of $(1-\delta^2)$.
Therefore, also the anomalous dimension of the nematic order parameter field $\eta_\varphi = 4 - d$ is expected to be exact at the AIAO fixed point.
The only exponent left to think about is the correlation-length exponent $\nu$.
Note that the AIAO and nematic propagator display the same exponent $\nu = \nu_\phi = \nu_\varphi$, since there is only one divergent length scale present at the AIAO fixed point, as demonstrated in the previous section.
Analogously to the flow equation of the anisotropy $\delta$, the effective charge $e^2$, and the AIAO Yukawa coupling $g_1^2$, higher-loop corrections to the flow of the AIAO mass $r_1$ are either strongly suppressed by strictly positive powers of $(1-\delta^2)$ and/or by strictly positive powers of $1/(1 + r_2)$ in the vicinity of the AIAO fixed point.
As a consequence, the stability matrix is of block triangular form with $1/\nu = 2 - \eta_\phi = d - 2$ as the only (strictly) positive eigenvalue.

\section{Further examples of continuous order-to-order transitions}
\label{app:further-examples}

In this appendix, we provide two further examples for continuous order-to-order transitions arising from fixed-point annihilation in interacting many-body systems.

\subsection{Anisotropic spin-boson model}

\begin{figure*}[t!]
\centering\includegraphics[width=0.84\linewidth]{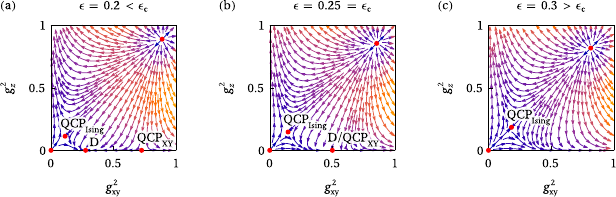}
\caption{%
(a)~RG flow diagram of the anisotropic spin-boson model for $\epsilon = 0.2$ in the plane spanned by $g_{xy}^2$ and $g_z^2$.
Arrows indicate flow towards infrared.
For $g_{xy}^2 < g_z^2$ at the ultraviolet scale, the parameters flow towards $(g_{xy}^2, g_z^2) \to (0,+\infty)$, associated with an Ising ordered phase.
For $g_{xy}^2$ above, but not too far from $g_{z}^2$, the flow is attracted by the fully infrared stable fixed point D, corresponding to an interacting disordered phase.
For large $g_{xy}^2$,  the parameters flow towards $(g_{xy}^2, g_z^2) \to (+\infty,0)$, associated with an XY ordered phase.
(b)~Same as (a), but for $\epsilon = \epsilon_\mathrm{c} = 1/4$, illustrating the collision of the quantum critical fixed point QCP$_\text{XY}$ and the infrared stable fixed point D.
(c)~Same as (a), but for $\epsilon = 0.3$. For all values of $g_{xy}^2 > g_z^2$, the parameters flow towards $(g_{xy}^2, g_z^2) \to (+\infty,0)$, associated with the $xy$-localized phase. The quantum critical fixed point QCP$_\text{Ising}$ on the isotropic axis $g_{xy}^2 = g_z^2$ governs a continuous transition between the two different orders.
}
\label{fig:flow-diagram-Bose-Kondo}
\end{figure*}

As an example for a model featuring a continuous order-to-order transition from fixed-point annihilation that is directly amenable to large-scale numerical simulations, consider the anisotropic spin-boson model, which describes a single spin $\vec S$ in a fluctuating magnetic field $\vec h$ with an XXZ anisotropy, defined by the Hamiltonian~\cite{sengupta00}
\begin{align}
	\mathcal{H}_\text{spin-boson} = g_{xy} (h^x S^x + h^y S^y) + g_z h^z S^z + \mathcal H_\text{bulk}(\vec h).
\end{align}
Here, $g_{xy}$ and $g_z$ correspond to in-plane and out-of-plane components of the $g$ tensor, respectively,
and the bulk Hamiltonian $\mathcal H_\text{bulk}(\vec h)$ is chosen such that the fluctuations of the magnetic field are Gaussian, with power-law correlation
\begin{align}
	\langle \mathcal{T}_\tau h^a (\tau) h^b (0) \rangle \propto \frac{\delta^{ab}}{| \tau |^{2 - \epsilon}},
\end{align}
where $\mathcal T_\tau$ corresponds to the time ordering operator in imaginary time $\tau$. The bath exponent $\epsilon$ represents an external tuning parameter, and we consider the case of small positive $\epsilon \in (0, 1/2)$.
The above spin-boson model may be realized as an effective description of a quantum impurity in a two-dimensional antiferromagnet in the vicinity of a quantum critical point~\cite{sachdev99,sachdev04,vojta06}.

The RG flow equations for the two couplings $g_{xy}$ and $g_z$ up to two-loop order read~\cite{zhu02,zarand02}
\begin{align}
    \frac{\rmd g^2_{xy}}{\rmd \ln b} & = \epsilon g^2_{xy} - g^2_{xy} (g^2_{xy} + g^2_z)
    + g_{xy}^4 (g^2_{xy} + g^2_z),
    \allowdisplaybreaks[1] \\
    \frac{\rmd g^2_z}{\rmd \ln b} & = \epsilon g^2_z - 2 g^2_{xy} g^2_z + 2 g^2_{xy} g_z^4,
\end{align}
where we have rescaled $g^2_{xy} \mapsto \Lambda^\epsilon g^2_{xy}$ and $g_z^2 \mapsto \Lambda^\epsilon g_z^2$.
The flow is illustrated for different values of $0< \epsilon < 1/2$ in Fig.~\ref{fig:flow-diagram-Bose-Kondo}.
For $\epsilon>0$, both couplings $g_{xy}^2$ and $g_z^2$ are infrared relevant.
For $g_{xy}^2 < g_z^2$ at the ultraviolet scale, the parameters flow towards $(g_{xy}^2, g_z^2) \to (0,+\infty)$. This regime is associated with an Ising ordered phase, in which the spin forms local order with a finite moment along the $z$ direction. The $z$-localized phase spontaneously breaks Ising $\mathbb Z_2$ symmetry~\cite{weber24}.
For $g_{xy}^2 > g_z^2$ at the ultraviolet scale, the phase structure depends on the value of the external tuning parameter $\epsilon$. This is due to a fixed-point annihilation that takes place at a critical value of the tuning parameter $\epsilon_\mathrm{c} = 1/4$. 
For $\epsilon < \epsilon_\mathrm{c} = 1/4$ and $g_{xy}^2$ above, but not too far from $g_z^2$, the flow is attracted by a fully infrared stable fixed point, corresponding to an interacting disordered phase (D). For large $g_{xy}^2$, the parameters flow towards $(g_{xy}^2, g_z^2) \to (+\infty,0)$. This regime is associated with an XY ordered phase, in which the spin forms local order with a finite moment perpendicular to the $z$ direction. 
For $\epsilon = \epsilon_\mathrm{c}$, the fixed point D collides with the quantum critical fixed point associated with the onset of XY order.
Now for $\epsilon > \epsilon_\mathrm{c}$, the parameters flow towards $(g_{xy}^2, g_z^2) \to (+\infty,0)$ for all values of $g_{xy}^2 > g_z^2$.
Most importantly, on the isotropic axis $g_{xy}^2 = g_z^2$, there is a quantum critical fixed point QCP$_\text{Ising}$ that governs a continuous transition between Ising order and the disordered phase for $0 < \epsilon < \epsilon_\mathrm{c}$ and between the Ising order and XY order phase for $\epsilon_\mathrm{c} < \epsilon < 1/2$.
For $\epsilon$ above, but not too far from $\epsilon_\mathrm{c}$, the two-loop flow equations thus suggest a continuous order-to-order transition governed by the QCP$_\text{Ising}$ fixed point.
In fact, the qualitative picture indicated by the perturbative calculation has recently been confirmed in large-scale numerical calculations~\cite{weber23, weber24}; see also Refs.~\cite{nahum22, cuomo22} for recent analytical approaches based on large-$S$ expansions.
The numerical calculations show that the true critical value of the control parameter $\epsilon$, at which the quantum critical fixed point QCP$_\text{XY}$ and the infrared stable fixed point D collide, is $\epsilon_\mathrm{c} = 0.2294(1)$, and thus close to our perturbative estimate.
Interestingly, the numerical data for the transition between the Ising and XY ordered phases (see Fig.~9(d) of Ref.~\cite{weber24}) confirm our theoretical predictions for a continuous order-to-order transition from fixed-point annihilation, as formulated in the main text: First, on the XY side of the transition, the corresponding order parameter is significantly suppressed in comparison with its counterpart on the Ising side. Second, finite-temperature effects are significantly more enhanced on the XY side of the transition. Both properties can be understood as direct consequence of the annihilation between the QCP$_\text{XY}$ and D fixed points and the resulting slow flow on the XY side of the transition.

\subsection{Quantum chromodynamics with four-fermion interactions}

\begin{figure*}[t!]
\includegraphics[width=\linewidth]{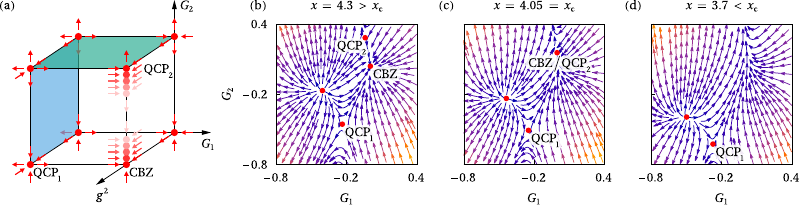}
\caption{%
(a)~Schematic fixed-point structure of QCD$_4$ with additional four-fermion interactions in the Veneziano limit $N_\text{flavor}, N_\text{color} \to \infty$, with fixed ratio $x = N_\text{flavor} / N_\text{color} < 11/2$, in the space spanned by the couplings $g^2$, $G_1$, and $G_2$.
Arrows indicate flow towards infrared.
For $x$ below, but not too far from $11/2$, the infrared attractive plane $g^2 = g^2_\star(G_1) = (11 - 2 x - 3 x G_1)/(13 x - 34)$ hosts a fully infrared stable Caswell-Banks-Zaks fixed point (CBZ), as well as two quantum critical fixed points (QCP$_1$ and QCP$_2$).
The corresponding critical manifolds are indicated in blue and green, respectively.
For decreasing $x$, CBZ and QCP$_2$ approach each other and eventually collide at a critical $x_\text{c}$.
(b)~RG flow diagram for $x=4.3$ in the plane $g^2 = g^2_\star$ spanned by the four-fermion couplings $G_1$ and $G_2$, indicating the locations of the infrared stable CBZ fixed point and the quantum critical fixed points QCP$_1$ and QCP$_2$.
(c)~Same as (b), but for the critical ratio $x = x_\text{c} = 4.05$, indicating the fixed-point collision.
(d)~Same as (b), but for $x=3.7$. The absence of the stable CBZ fixed point implies a direct continuous transition between a chiral-symmetry-broken state, signaled by the runaway flow towards $G_2 \to +\infty$, and a different ordered ground state, signaled by the runaway flow towards $(G_1, G_2) \to (-\infty, -\infty)$.
}
\label{fig:flow-diagram-QCD4}
\end{figure*}

As an example for a continuous order-to-order transition relevant to high-energy physics, we discuss quantum chromodynamics in $3+1$ space-time dimensions (QCD$_4$), supplemented with local four-fermion interactions, defined by the action~\cite{gies04b,gies06,braun11,kusafuka11,gukov17,kuipers19}
\begin{align}
    S_\text{QCD$_4$-GN} = 
    \int \dd[4]{x} \left[ \rmi \bar\psi \slashed{D} \psi + \frac{1}{4} F_z^{\mu\nu} F_{\mu\nu}^z + \frac{1}{2} \sum_{\alpha=1}^4 G_\alpha \mathcal{O}_\alpha \right]
\end{align}
with covariant derivative $\slashed{D} \coloneqq \gamma_\mu (\partial_\mu - \rmi g A_\mu)$,
SU($N_\text{color}$) gauge field $A_\mu \coloneqq A_\mu^z T^z$, and the associated field-strength tensor $F_{\mu\nu} \coloneqq F_{\mu\nu}^z T^z$, where $T^z$ denote the generators of SU$(N_\text{color})$.
A Fierz-complete basis of local four-fermion interactions is given by~\cite{gies04b}
\begin{align}
\mathcal{O}_1 & \coloneqq \left( \bar\psi^a \gamma_\mu \psi^b \right)^2 + \left( \bar\psi^a \gamma_\mu \gamma_5 \psi^b \right)^2, \\
\mathcal{O}_2 & \coloneqq \left( \bar\psi^a \psi^b \right)^2 - \left( \bar\psi^a \gamma_5 \psi^b \right)^2, \\
\mathcal{O}_{3,4} & \coloneqq \left( \bar\psi \gamma_\mu \psi \right)^2 \pm \left( \bar\psi \gamma_\mu \gamma_5 \psi \right)^2.
\end{align}
For notational simplicity, we have abbreviated $\bar\psi \psi \equiv \bar\psi_i^a \psi_i^a$ and $(\bar\psi^a \psi^b)^2 \equiv \bar\psi^a_i \psi^b_i \bar\psi^b_j \psi^a_j$, and assumed summation convention over repeated color indices $i,j$ and flavor indices $a,b$.
Consider the so-called Veneziano limit $N_\text{flavor}, N_\text{color} \rightarrow \infty$ with fixed ratio $x \coloneqq N_\text{flavor} / N_\text{color}$.
In this limit, the flow equations reduce to a set of three equations~\cite{kusafuka11,gukov17,kuipers19},
\begin{align}
    \frac{\rmd g^2}{\rmd \ln b} & = \frac{2 (11 - 2 x)}{3} g^4 + \frac{2 (34 - 13 x)}{3} g^6
    \nonumber\\&\quad
    - 2 x g^4 G_1,
    \label{eq:flow-gauge-coupling-QCD}\allowdisplaybreaks[1] \\
    \frac{\rmd G_1}{\rmd \ln b} & = -2 G_1 - (1 + x) G_1^2 - \frac{x}{4} G_2^2 + \frac{3}{4} g^4,
    \allowdisplaybreaks[1] \\
    \frac{\rmd G_2}{\rmd \ln b} & = -2 G_2 + 2 G_2^2 - 2 x G_1 G_2 + 6 g^2 G_2 + \frac{9}{2} g^4,
\end{align}
where we have rescaled the gauge coupling $g^2 N_\text{color} / (16 \pi^2) \mapsto g^2$ and the four-fermion couplings $G_{1,2} \Lambda^2 N_\text{color} / (4 \pi^2) \mapsto G_{1,2}$.
For reasons of consistency with the previous sections, we write the flow equations in terms of an RG scale $\ln b$, with $\ln b = 0$ ($\ln b \to \infty$) corresponding to the ultraviolet (infrared) scale.

The fixed-point structure of the model in the Veneziano limit is shown schematically in Fig.~\ref{fig:flow-diagram-QCD4}(a).
For $x > 11/2$, asymptotic freedom is lost and the gauge coupling vanishes in the infrared if $G_1$ and $G_2$ are small at the ultraviolet scale.
For $x < 11/2$, the gauge charge is a RG relevant parameter and flows towards a finite value $g^2 = g^2_\star(G_1) = ({11 - 2 x - 3 x G_1})/({13 x - 34})$ in the infrared.
For $x$ below, but not too far from $11/2$, the infrared attractive plane $g^2 = g^2_\star(G_1)$ hosts a fully infrared stable fixed point that can be associated with the conformal phase of many-flavor QCD$_4$, the so-called Caswell-Banks-Zaks fixed point (CBZ)~\cite{caswell74,banks82}.
In addition, the plane features two fixed points with unique infrared relevant directions, which can be associated to quantum critical points (QCP$_1$ and QCP$_2$).
For decreasing $x$, the infrared stable CBZ fixed point and one of the two quantum critical points (QCP$_2$) approach each other and eventually collide at the critical value $x_\text{c} = 4.05$.
This is illustrated in Figs.~\ref{fig:flow-diagram-QCD4}(b)--(d), which show the RG flow in the plane $g^2 = g^2_\star(G_1)$ for different values of $x$.
The runaway flow towards $G_2 \to +\infty$ at finite RG scale is usually associated with an instability towards a chiral symmetry breaking ground state~\cite{gies06,kaplan09,braun11}.
The other quantum critical point (QCP$_1$) in the infrared attractive plane $g^2 = g^2_\star(G_1)$, however, is not affected by the fixed-point annihilation and persists for all $x > 0$.
Since the runaway flow corresponding to this quantum critical fixed point is towards $(G_1,G_2) \to (-\infty, - \infty)$, one should expect it to describe an instability towards a different ground state, spontaneously breaking a symmetry different from chiral symmetry.
This suggests a continuous order-to-order transition for $x < x_\text{c}$ governed by QCP$_1$.

\small

\bibliographystyle{longapsrev4-2}
\bibliography{QBT3D-nem-Weyl}

\end{document}